\documentclass{article}

\usepackage{arxiv}          

\usepackage{xcolor}
\definecolor{darkred}{rgb}{0.65,0.0,0.0}
\definecolor{darkblue}{rgb}{0.0,0.0,0.5}

\usepackage[utf8]{inputenc} 
\usepackage[T1]{fontenc}    
\usepackage{hyperref}       
\hypersetup{                
    colorlinks=true,
    linkcolor=darkblue,
    filecolor=darkblue,
    urlcolor=darkblue,
    citecolor=darkblue
    }

\usepackage{url}            
\usepackage{booktabs}       
\usepackage{amsfonts}       
\usepackage{nicefrac}       
\usepackage{microtype}      

\usepackage{graphicx}
\graphicspath{{figures/}}   

\usepackage[numbers]{natbib}

\usepackage{doi}            

\usepackage{caption}
\usepackage{subcaption}
\usepackage{chemist}
\usepackage{chemformula}
\usepackage{color, colortbl,booktabs}
\definecolor{Gray}{gray}{0.95}

\usepackage{siunitx}
\usepackage{authblk}

\author[a]{{Tássia L.S. Quaresma }}
\author[b]{{Tristan Hehnen }}
\author[a,b]{{Lukas Arnold }}

\affil[a]{Institute for Advanced Simulation, Forschungszentrum  J\"ulich, Wilhelm-Johnen-Straße, 52428 J\"ulich, Germany}
\affil[b]{Chair of Computational Civil Engineering, University of Wuppertal, Pauluskirchstraße 7, 42285 Wuppertal, Germany\newline}

\affil[ ]{
Tássia L.S. Quaresma: \texttt{\href{mailto:t.quaresma@fz-juelich.de}{t.quaresma@fz-juelich.de}; \href{https://orcid.org/0000-0003-0214-0595}{ORCID: 0000-0003-0214-0595}}
}

\affil[ ]{
Tristan Hehnen: \texttt{\href{mailto:hehnen@uni-wuppertal.de}{hehnen@uni-wuppertal.de}; \href{https://orcid.org/0000-0002-6123-261X}{ORCID: 0000-0002-6123-261X}}
}

\affil[ ]{
Lukas Arnold: \texttt{\href{mailto:l.arnold@fz-juelich.de}{l.arnold@fz-juelich.de}; \href{mailto:arnold@uni-wuppertal.de}{arnold@uni-wuppertal.de}; \href{https://orcid.org/0000-0002-5939-8995}{ORCID: 0000-0002-5939-8995}}
}

\title{Sensitivity Analysis for an Effective Transfer of Estimated Material Properties from Cone Calorimeter to Horizontal Flame Spread Simulations}
\renewcommand{\shorttitle}{Sensitivity Analysis for an Effective Transfer of Estimated Material Properties }

\hypersetup{
pdftitle={A hot fire experiment with a cool modelling approach},
pdfsubject={Fire Propagation Simulation},
pdfauthor={First Author, Second Author, Third Author},
pdfkeywords={fire, experiment, model, more keywords},
}

\begin{document}
\maketitle

\begin{abstract}
Predictive flame spread models based on temperature dependent pyrolysis rates require numerous material properties as input parameters. These parameters are typically derived by optimisation and inverse modelling using data from bench scale experiments such as the cone calorimeter. The estimated parameters are then transferred to flame spread simulations, where self-sustained propagation is expected. A fundamental requirement for this transfer is that the simulation model used in the optimisation is sufficiently sensitive to the input parameters that are important to flame spread. Otherwise, the estimated parameters will have an increased associated uncertainty that will be transferred to the flame spread simulation. This is investigated here using a variance-based global sensitivity analysis method, the Sobol indices. The sensitivities of a cone calorimeter and a horizontal flame spread simulation to 15 effective properties of polymethyl methacrylate are compared. The results show significant differences between the setups: the cone calorimeter is dominated by strong interaction effects between two temperature dependent specific heat values, whereas the flame spread is influenced by several parameters. Furthermore, the importance of some parameters for the cone calorimeter is found to be time-varying, suggesting that single-value cost functions may not be sufficient to account for all sensitive parameters during optimisation.

\end{abstract}

\keywords{global sensitivity analysis \and Sobol indices \and horizontal flame spread simulation \and Cone Calorimeter simulation \and polymethyl methacrylate (PMMA) \and material property estimation \and inverse modelling \and optimisation \and Fire Dynamics Simulator (FDS)}

\section{Introduction}
\label{sec:intro}

The development of fire simulation models that are capable of predicting the behaviour of flame spread from material properties is essential to overcome limitations of prescriptive design fires. By accounting for the heat transfer inside the solid and the coupling between the pyrolysis rates and the material temperatures, a responsive heat release rate (HRR) to changes in the surrounding conditions can be predicted, rather than prescribed. However, one major constraint in the development of predictive models is the difficulty in experimentally measuring all the required material properties, which are taken as model input parameters. As an alternative strategy, optimisation algorithms have been applied to parameter estimation in a so-called inverse modelling process (IMP)~\cite{hehnen2022PMMA, viitanen2022cfd, fiola2021comparison, hehnen2020numerical, yang2019methodology, nyazika2019pyrolysis, beji2019numerical, arnold2018propti, hostikka2017pyrolysis, kim2015parameter}. In the IMP, data from bench-scale experiments are used as a target for determining the set of input parameters that lead to the closest fit between the simulation output and the experimental data. Typically, the HRR measured from Cone Calorimeter experiments is taken as target, and the fitness is evaluated by a cost function. The performance of the estimated parameter set is subsequently validated in simulation setups of different scales, where a self-sustained flame spread is expected to occur. 


This approach assumes that the sensitivities of the model output to the input parameters are equivalent, both in the bench-scale setup and in setups involving the flame spread. However, this assumption might not hold, particularly when the estimated parameter set is to be validated in flame spread simulations of increased scales. One possible reason is that the bench scale experiments are performed under conditions (e.g. small sample size, uniform heating) where the flame spread is either negligible or simply does not occur. In addition, the bench scale experiments are difficult to be modelled sufficiently well because of unknown boundary conditions, limitations of existing sub-models, and difficulties in achieving high grid resolutions due to the prohibitive computational cost. As a consequence, it is not possible to affirm that an estimated parameter set, that performs well in the Cone Calorimeter, will achieve an equivalent satisfactory result when transferred to the flame spread simulation~\cite{hehnen2022PMMA, beji2019numerical}.

In this contribution, we focus on answering the following question: how sensitive is the Cone Calorimeter simulation setup to the parameters that are important to the flame spread simulation? This is crucial because parameters with low importance to the Cone Calorimeter simulation will be estimated with an enlarged degree of uncertainty, which is then carried over to the flame spread simulation. This happens because their effect on the cost function used in the optimisation is expected to be comparably low, leading the optimiser to freely choose any value from the pre-defined sampling ranges. If such parameters, on the other hand, are important to the flame spread simulation, their uncertainty is propagated, which can compromise the reliability of the simulated results.

In this regard, we carry out sensitivity analyses (SAs) on two simplified simulation setups: a Cone Calorimeter; and a horizontal flame spread, to evaluate the models' sensitivities to a set of 15 input parameters. The parameter set consists of effective material properties of polymethyl methacrylate (PMMA), of its pyrolysis residue, and of an insulation material. The parameters were estimated in an IMP that used the same Cone Calorimeter simulation setup investigated here. Both, the set of input parameters and the simplified Cone Calorimeter simulation, were proposed and validated in a previous study by  Hehnen and Arnold~\cite{hehnen2022PMMA}. The flame spread setup represents a generic horizontal configuration, in which a self-sustained spread develops on a PMMA sample in still air. To the author's best knowledge, only a limited number of studies have investigated the flame spread behaviour on PMMA samples in such configuration, either experimentally~\cite{korobeinichev2018experimental, jiang2017sample} or numerically~\cite{karpov2018numerical}. For this reason, this configuration is specifically chosen. All the simulations discussed in the present work were conducted with a self-compiled version of the Fire Dynamics Simulator (FDS, version FDS6.7.6-810-ge59f90f-HEAD)~\cite{mcgrattan2005fire}.


Past studies~\cite{fleurotte2022sensitivity, nyazika2019pyrolysis, yang2019methodology, batiot2016sensitivity, kim2015parameter} have addressed the importance of running SAs in order to improve the efficiency of strategies for material property estimation, which are based on multi-objective optimisation. In general, the SA is aimed for model simplification, by determining the relative importance of input parameters, such that the unimportant ones can be filtered out from the optimisation, saving computing time. More recently, Ding~et~al.~\cite{ding2023analysis} conducted one-at-a-time SAs on large-scale upward flame spread simulations to identify which input parameters affect the spread the most, so that their measurement/ estimation can be improved. Here, we aim at filling this gap, by determining not only the most influential parameters to the flame spread, but specifically at identifying whether parameter importance changes from one setup (Cone Calorimeter) to another (horizontal flame spread). 


The SAs performed in this contribution are mainly discussed in terms of the Sobol indices~\cite{sobol2001global, saltelli2002making}, a robust global SA method based on the decomposition of variances. By varying input parameters simultaneously and not one at a time, the method is capable of quantitatively capturing interaction effects between input parameters on the model output of interest. It is therefore suitable for determining the sensitivities of non-linear and high-dimensional models, such as the complex pyrolysis models in question. Sensitivities are determined based on the degree of contribution that a certain input parameter has to the uncertainty (variance) of the model output.  It has been applied in the field of fire safety science to investigate the influence of inputs on environmental fire spread models~\cite{ujjwal2021global}, and on the mass loss rate calculated by the Arrhenius equation~\cite{batiot2016sensitivity}. In this contribution, the Sobol indices are estimated to express the effects of the input parameters on different simulation outputs: 

\begin{itemize}
    \item the temporal development of the HRR, which is convenient to assess how the influence of a certain input parameter varies throughout the course of the simulation;
    \item the root mean square error (RMSE), calculated between the simulated HRRs and the measured HRR, commonly used as a cost function in the IMP;
    \item and the mean rate of spread (MRS), calculated in an additional post-processing step based on the derivative of the flame front position with respect to time.
\end{itemize}


In addition, the relation between the RMSE and the MRS and their two respective most influential input parameters are qualitatively discussed on the basis of scatter plots, providing a visual representation of their responses to changes in the inputs.

This article is accompanied by a publicly available data repository on Zenodo \cite{zenodo:ArticleDataset}, containing the simulation data, the Python scripts used for data analysis, and supplementary material data.

\section{Methods}
\label{sec:methods}

\subsection{Cone Calorimeter simulation}\label{CC_setup}

The simulation setup of a simplified Cone Calorimeter considered as reference case in this work stems from freely available previous studies~\cite{hehnen2022PMMA}. The model was initially developed to be used in an IMP for estimating thermophysical properties of PMMA, based on data of Cone Calorimeter experiments of black cast PMMA.  The simulations were conducted with the same version used in the scope of this work for consistency. The experimental data was provided by the Aalto University to the publicly available MaCFP database~\cite{macfp2022material}. In the Cone Calorimeter experiments, a square sample of PMMA with 10~cm edge length and 6~mm thickness is exposed to a radiative heat flux of 65~kW/m$^2$. The insulation material of equal surface area is positioned below the PMMA sample for insulation, and is 2~cm thick.

\begin{figure}[!htp]
     \centering
     \begin{subfigure}[b]{0.49\textwidth}
         \centering
          \includegraphics[width=\textwidth]{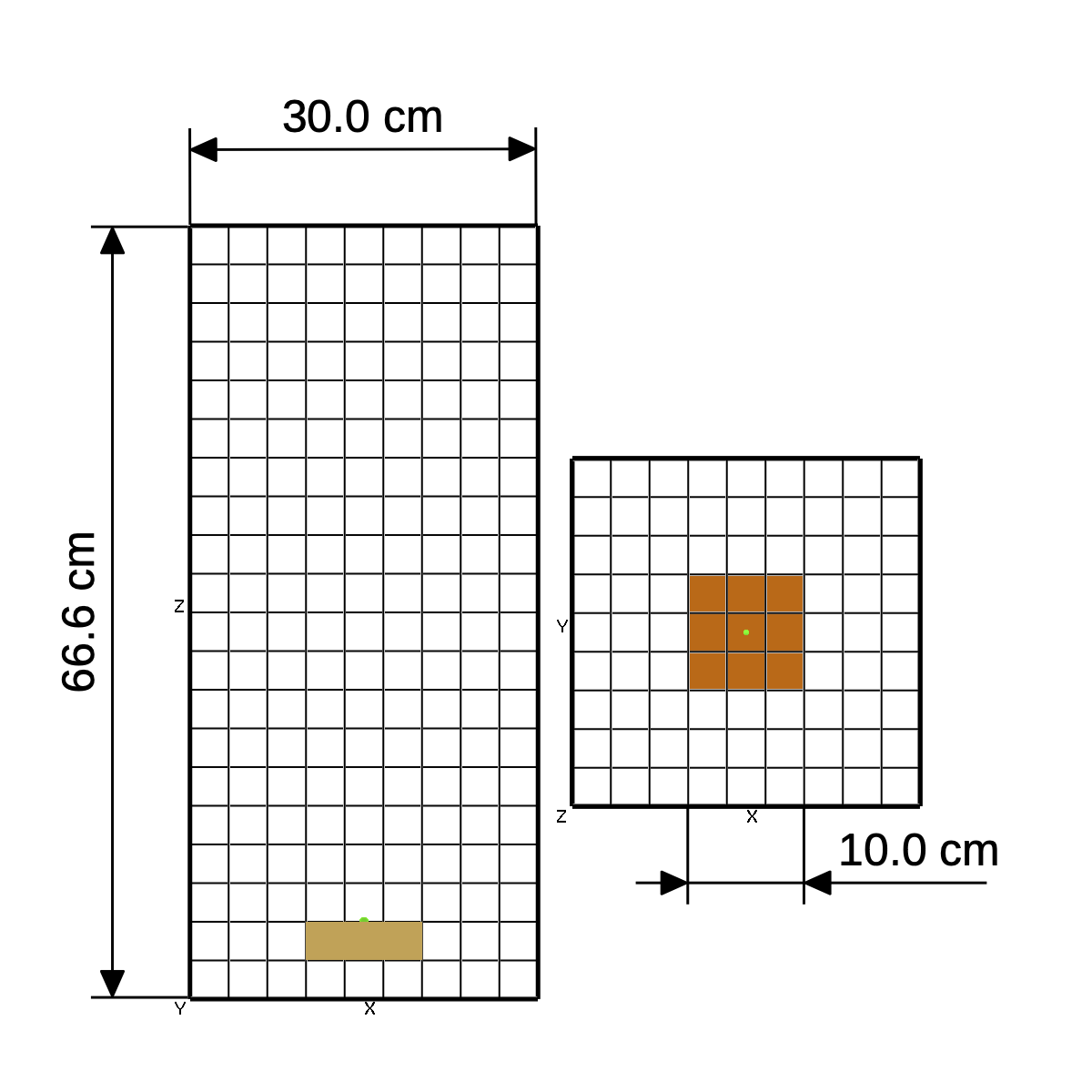}
         \caption{Dimensions and fluid phase resolution.}
         \label{cone04a}
     \end{subfigure}
     \hfill
     \begin{subfigure}[b]{0.49\textwidth}
         \centering
         \includegraphics[width=\textwidth]{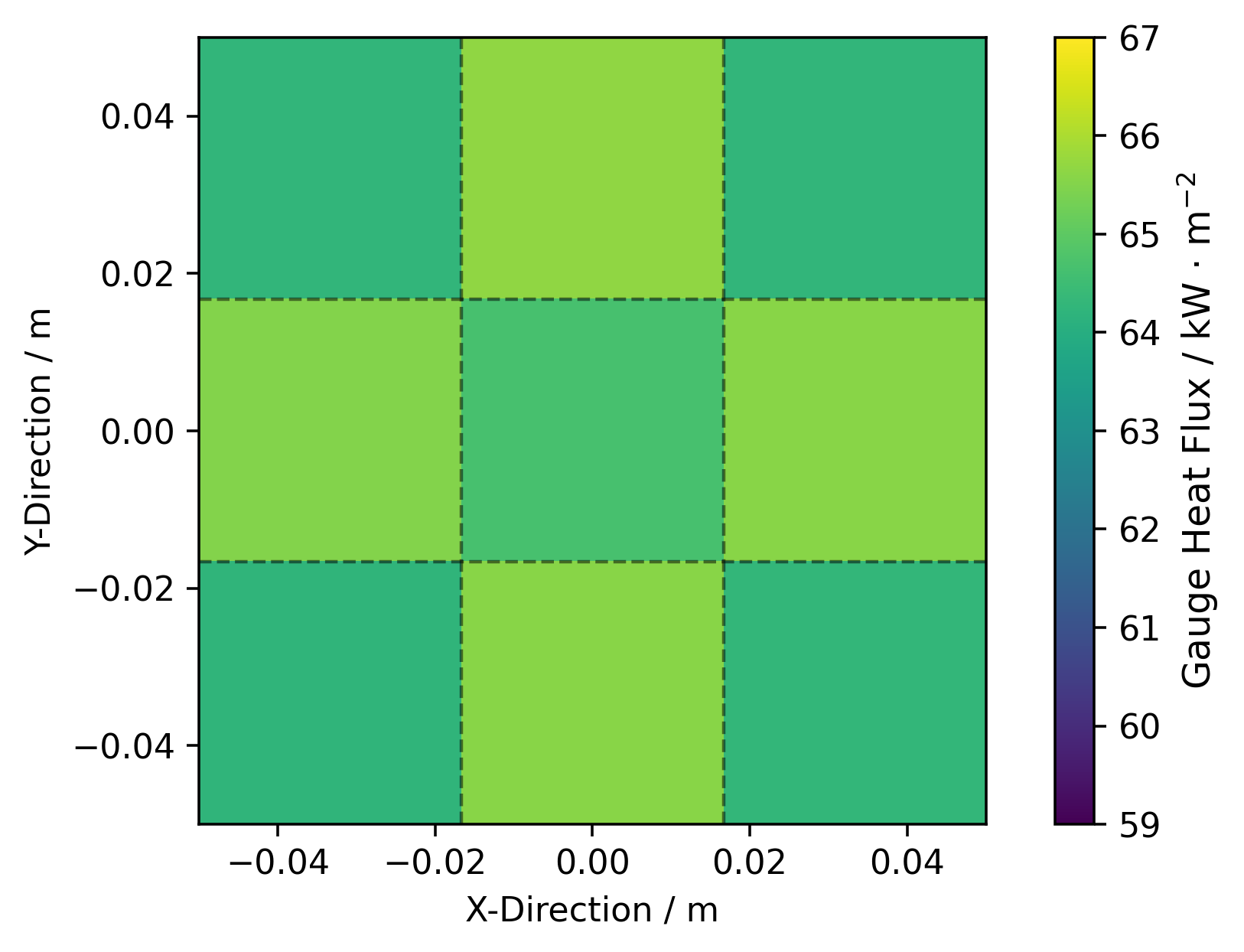}
         \caption{Gauge heat flux mapping applied to the sample surface.}
         \label{cone04b}
     \end{subfigure}
     \hfill
     \begin{subfigure}[b]{0.49\textwidth}
         \centering
         \includegraphics[width=\textwidth]{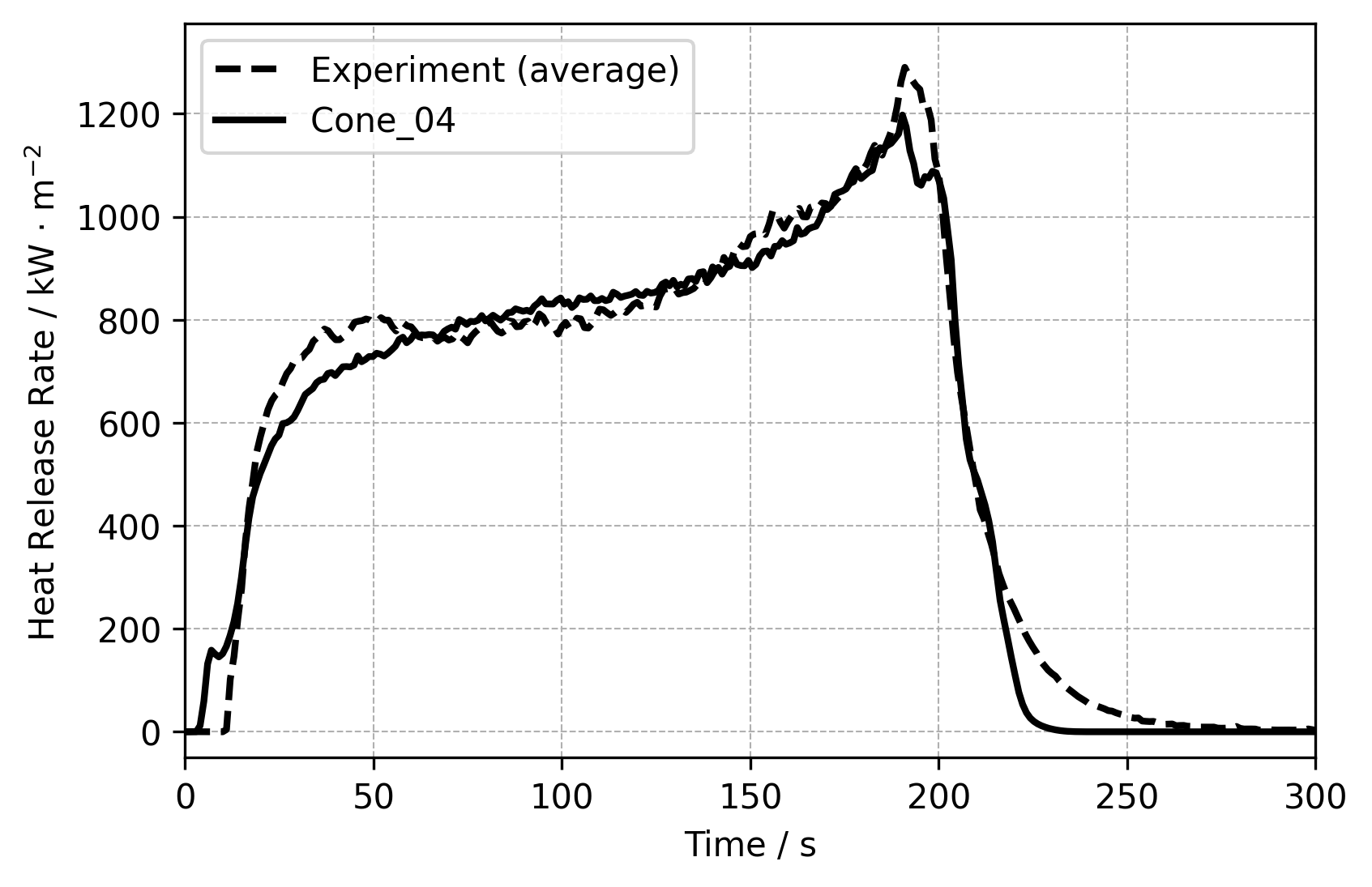}
         \caption{Best fit of simulated HRR to experimental data.}
         \label{cone04c}
     \end{subfigure}
        \caption{Simulation setup of the simplified Cone Calorimeter, ``Cone\_04", used for estimating thermophysical parameters of black cast PMMA \cite{hehnen2022PMMA}, taken as reference case. }
        \label{cone04}
\end{figure}

Amongst the different Cone Calorimeter simulation models presented previously~\cite{hehnen2022PMMA}, we chose the one labelled as ``Cone\_04". Cone\_04 uses an increased resolution when compared to similar approaches in the field, in which Cone Calorimeter models were also employed to estimate material properties. Viitanen et al. \cite{viitanen2022cfd}, Hehnen et al. \cite{hehnen2020numerical}, and Beji and Merci \cite{beji2019numerical} considered fluid cell sizes of 5~cm, whereas in Cone\_04, 3.33~cm fluid cells are defined, see Figure~\ref{cone04a}. 

In Cone\_04, the radiative heat flux of 65~kW/m$^2$ is assigned to the sample surface, thus avoiding the need for modelling the heater. Despite this simplification, the non-uniform heating of the sample surface induced by the conical heating element is accounted for. This is accomplished, by defining multiple surfaces with slightly different values of heat fluxes, as presented in Figure~\ref{cone04b}. In addition, the model stood out for providing a set of thermophysical properties, which lead to the best fit to the experimental HRR data in the IMP, see Figure~\ref{cone04c}. For those reasons, Cone\_04 is expected to lead to an improved representation of the sample heating and thermal decomposition, and it is therefore taken here as reference case.

In the following subsections, we provide only a concise description of the model that is relevant to the scope of this contribution. For further details and additional resources, the reader should refer to the original work~\cite{hehnen2022PMMA} and to the FDS User's Guide~\cite{mcgrattan2005fire}, both freely available.

\subsubsection{Pyrolysis and solid phase} \label{pyro}

In this approach, it is assumed that PMMA pyrolysis can be described by an elementary first-order reaction of the form:

\begin{center}
    \ch{PMMA(solid) -> 0.99 Fuel Mixture(gas) + 0.01 Residue(solid)}    
\end{center}

where the fuel mixture is composed by methane, ethylene and carbon dioxide, and the residue is an inert solid product. The rates of pyrolysis are dependent on the local temperatures of the solid, and are calculated with the Arrhenius equation, as implemented in FDS~\cite{mcgrattan2005fire}. As a strategy to achieve a better fit to the experimental data, it was assumed that different fractions of the total PMMA mass decompose, each, at a different rate. The kinetic parameters (pre-exponential factor, activation energy), heats of reaction, and corresponding PMMA mass fractions were determined in a primary IMP step, taking data of MCC and TGA as target~\cite{hehnen2022PMMA}. In this same step, the volume fractions of the fuel mixture components were also determined. 

In this contribution, the kinetic parameters will not take part in the sensitivity analysis, given that our focus here is solely on the set of material properties that were estimated using the Cone Calorimeter. The parameter set of interest comprises only the thermophysical properties of PMMA, of the insulation material, and of the pyrolysis residue. For PMMA, these properties are: emissivity, absorption coefficient, refractive index, specific heat and thermal conductivity. The thermal conductivity and the specific heat of PMMA were both defined as temperature-dependent values, following a piecewise linear function with reference points at 150~°C, 480~°C and 800~°C. For the insulation material and the residue, the parameters are: emissivity, thermal conductivity and specific heat. In total, the material properties estimated in an IMP using the Cone Calorimeter count 15 input parameters. Their respective estimated values are presented in Table~\ref{parameters}.  

\begin{table}[ht]
    \centering
    \caption{Set of effective properties describing the solid phase materials. The parameters were estimated in the work of Hehnen and Arnold~\cite{hehnen2022PMMA} using the reference case Cone\_04.}
    \begin{tabular}{rllcc}
    \toprule
  & Material   &  Parameter & Estimated IMP value & Unit \\
    \midrule
1 & PMMA     & Emissivity              & 0.941    &  -        \\
2 &          & Absorption coefficient  & 7978.8   &  m$^{-1}$ \\
3 &          & Refractive index        & 2.854    &  - \\
4 &          & Conductivity at 150~°C   & 0.379  &  W$\cdot$m$^{-1}$$\cdot$K$^{-1}$ \\
5 &          & Conductivity at 480~°C   & 0.024   &  W$\cdot$m$^{-1}$$\cdot$K$^{-1}$ \\
6 &          & Conductivity at 800~°C   & 4.337   &  W$\cdot$m$^{-1}$$\cdot$K$^{-1}$ \\
7 &          & Specific heat at 150~°C  & 0.774   &  kJ$\cdot$kg$^{-1}$$\cdot$K$^{-1}$ \\
8 &          & Specific heat at 480~°C  & 3.808   &  kJ$\cdot$kg$^{-1}$$\cdot$K$^{-1}$ \\
9 &          & Specific heat at 800~°C  & 7.275   &  kJ$\cdot$kg$^{-1}$$\cdot$K$^{-1}$ \\
10 & Residue & Emissivity     & 0.552   &  - \\
11 &         & Conductivity   & 4.509   &  W$\cdot$m$^{-1}$$\cdot$K$^{-1}$  \\
12 &         & Specific heat  & 5.893   &  kJ$\cdot$kg$^{-1}$$\cdot$K$^{-1}$ \\
13 & Backing & Emissivity     & 0.441   &  - \\
14 &         & Conductivity   & 2.408   &  W$\cdot$m$^{-1}$$\cdot$K$^{-1}$   \\
15 &         & Specific heat  & 4.067  &  kJ$\cdot$kg$^{-1}$$\cdot$K$^{-1}$ \\
    \bottomrule
    \end{tabular}
    \label{parameters}
\end{table}

The value of PMMA density was directly calculated from reported mass and dimensions of the sample. Density of the residue was fixed to an arbitrary value due to lack of information, and density of the insulation material was taken from the MaCFP database~\cite{macfp2022material}.


The default one-dimensional heat conduction model in FDS was used~\cite{mcgrattan2005fire}. In this model, heat conduction is calculated only in the direction normal to the sample surface. The solid phase solution is updated at every time step and the node spacing of the PMMA layer is set to uniform. For the layer of the insulation material, the default stretched node spacing is considered. 

By default, the cell size of the solid phase is defined to be less than or equal to the square root of the thermal diffusivity. The default grid resolution in both layers is increased by a factor of 10, by setting the \verb|CELL_SIZE_FACTOR|~(CSF) to~0.10. This modification leads the PMMA layer to be discretised in 96 equally spaced cells, and the layer of the insulation material in 11 stretched cells. Nonetheless, the cell size of the PMMA layer is automatically re-defined during the simulation, as the temperature-dependant parameters affect the thermal diffusivity, as well as changing layer thickness due to sample consumption. 




\subsubsection{Combustion and gas phase} \label{comb}

Combustion of the fuel mixture is assumed mixing-controlled and the default simple chemistry model of FDS is used (Eddy Dissipation Concept). The soot yield is set to 0.022~g/g~\cite{quintiere1998principles}. The default grey gas model is used, considering a radiative fraction of 0.35 for the defined fuel mixture, which is the default value in FDS for unspecified species. The initial ambient temperature is set to 30.85~°C, consistent with what was reported from the experiment.

The Large Eddy Simulation (LES) is chosen as simulation mode. All default settings related to the models that accompany LES in FDS are kept unchanged. This means that sub-grid scales are modelled using the Deardorff model for the eddy viscosity, and the wall-adapting local eddy-viscosity (WALE) is used as near-wall turbulence model.

An overview of the geometry, domain and mesh resolution can be seen in Figure~\ref{cone04a}. The computational domain extends in the x- and y-directions from -15.0~cm to 15.0~cm, and in the z-direction from -6.6~cm to 60.0~cm. A uniform grid is defined by assigning 9~x~9~x~20~cells respectively in the x-, y-, z-directions, resulting in cells of 3.33~cm edge length. A single mesh is used for the whole domain, whose boundaries are set to open conditions. The centre of the sample is positioned at the origin (0,0,0). The sample holder is considered an inert obstruction of 10~cm edge length in the x- and y-directions and 3.33~cm in the z-direction.  



In order to evaluate grid independence, three additional simulation cases are built from the reference case Cone\_04, each considering a different fluid cell size. Following the terminology proposed earlier~\cite{hehnen2022PMMA}, Cone\_04 is here labelled as ``C3", as a reference to the 3x3 number of divisions of the sample surface, see Figures~\ref{cone04a} and~\ref{cone04b}. Thus, the first case is called~``C2" (2x2~divisions), referring to 5~cm cells, the second case is called~``C5" (5x5~divisions), with 2~cm cells in the fluid phase, and the third is~``C7" (7x7~divisions) with 1.43~cm cells. The heat flux mapping applied to the surface of the sample is adjusted in each case to conform with the C2, C5 and C7 resolutions. 

\begin{figure}[!htp]
         \centering
          \includegraphics[width=0.5\textwidth]{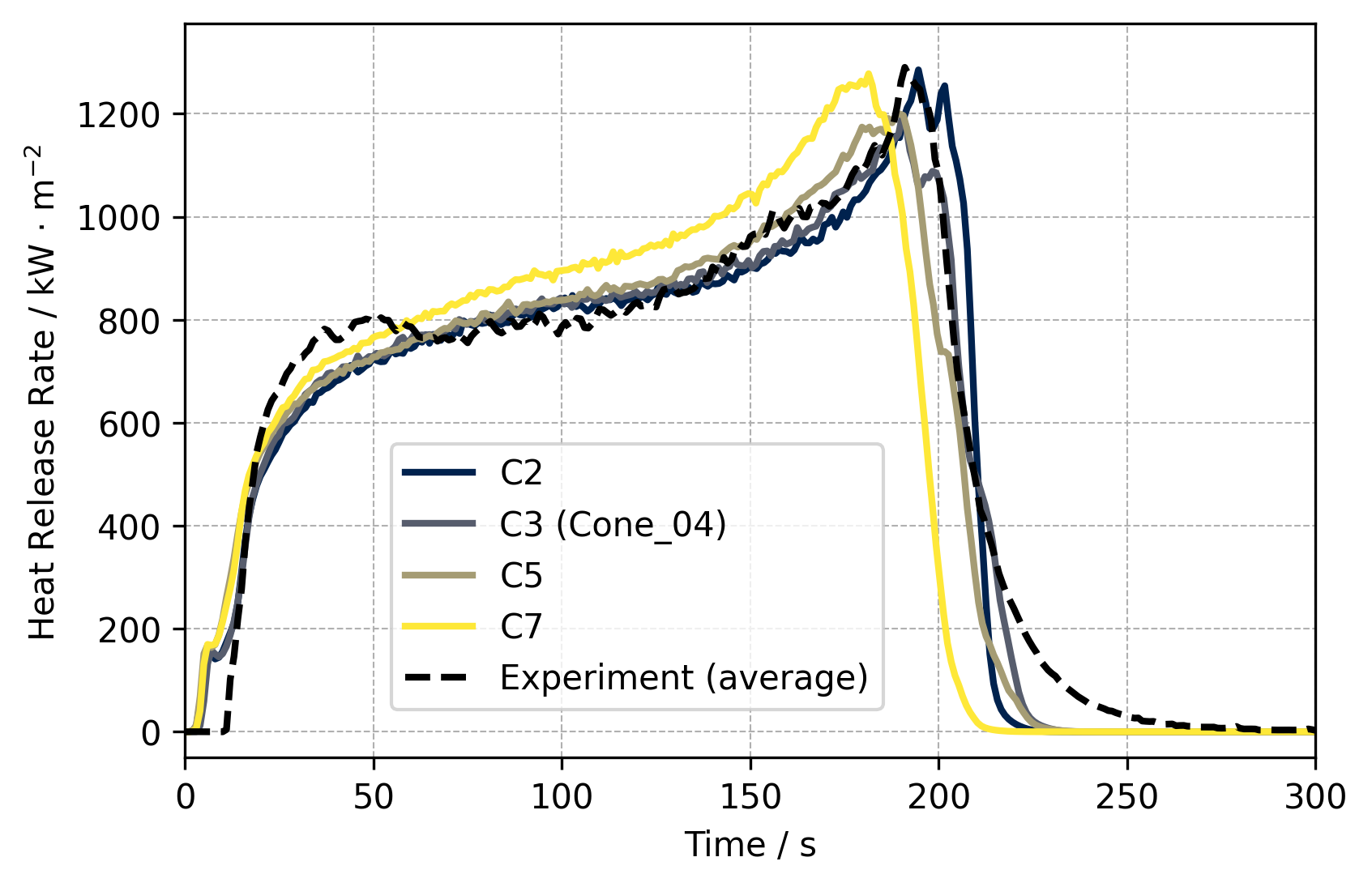}
        \caption{Effect of different fluid phase resolutions on the simulated HRR of the reference case Cone\_04.}
        \label{conegrid}
\end{figure}

From Figure~\ref{conegrid}, it is possible to observe that the simulated HRR does not significantly change with grid refinement in the fluid phase. The overall shape of the curve is maintained, and the main impact is on the region of the HRR peak, which is only slightly decreased and shifted to the left, as the fluid phase resolution increases. It is therefore expected, that the relative importance of the input parameters to the simulated HRR is conserved across different fluid cell sizes.

\subsection{Horizontal flame spread simulation} \label{fssetup}

The flame spread simulation considered in the SA represents a generic horizontal configuration, in which a self-sustained spread develops over a slab of PMMA. No influence of ventilation or wind conditions is considered, such that the flame is not disturbed by any external changes in the flow field.

The solid phase modelling used in the flame spread simulation is transferred from the reference case of the Cone Calorimeter simulation. This means that the PMMA sample, pyrolysis, the insulation material, and the solid phase resolution are the same as described in section~\ref{pyro}. This is important since the main goal of this contribution is to compare the responses of both simulations to variations in the same set of material properties presented in Table~\ref{parameters}.

\begin{figure}[!htp]
     \centering
     \begin{subfigure}[b]{0.49\textwidth}
         \centering
          \includegraphics[width=0.9\textwidth]{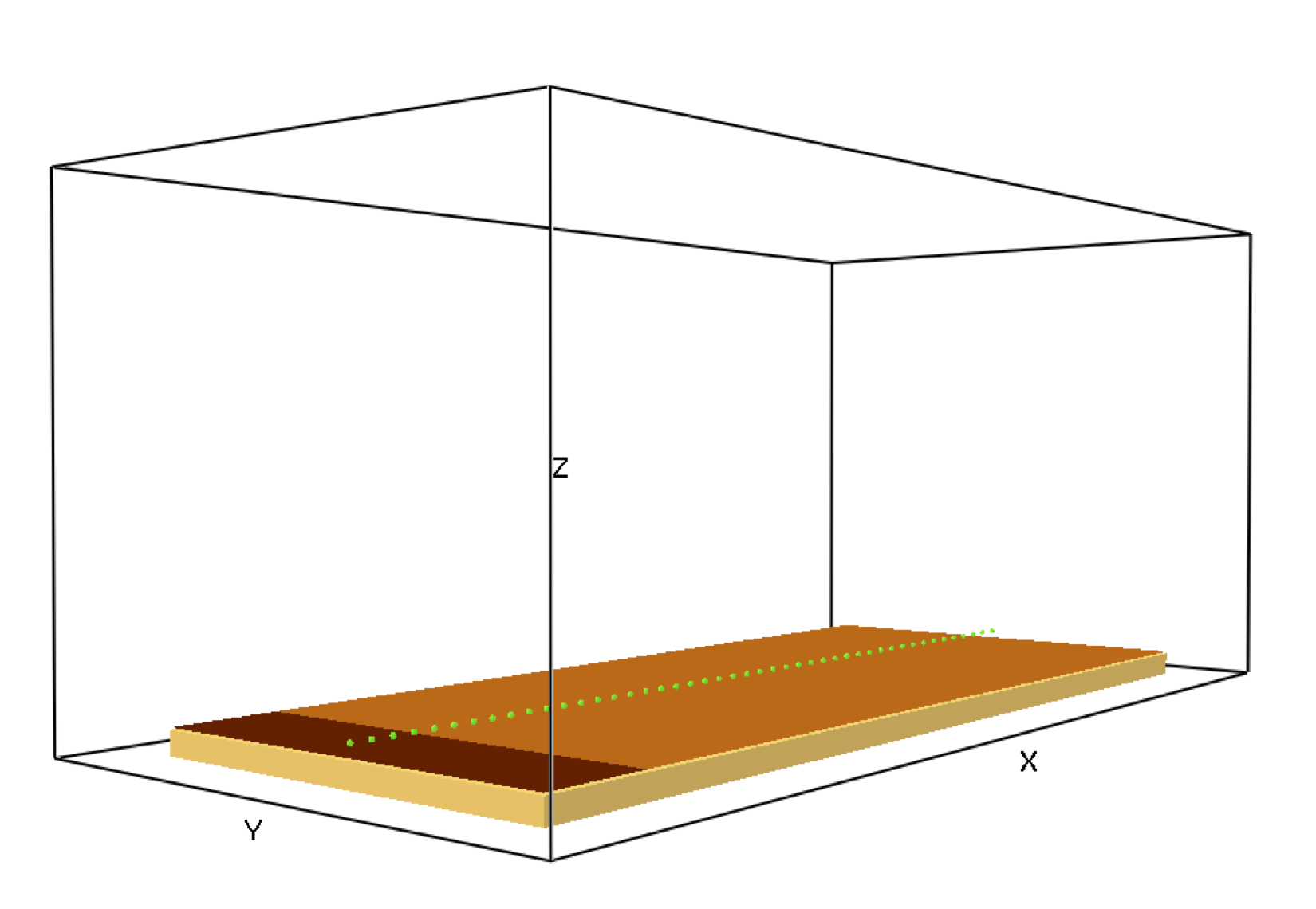}
         \caption{Perspective view.}
         \label{flamespreada}
     \end{subfigure}
     \hfill
     \begin{subfigure}[b]{0.49\textwidth}
         \centering
         \includegraphics[width=\textwidth]{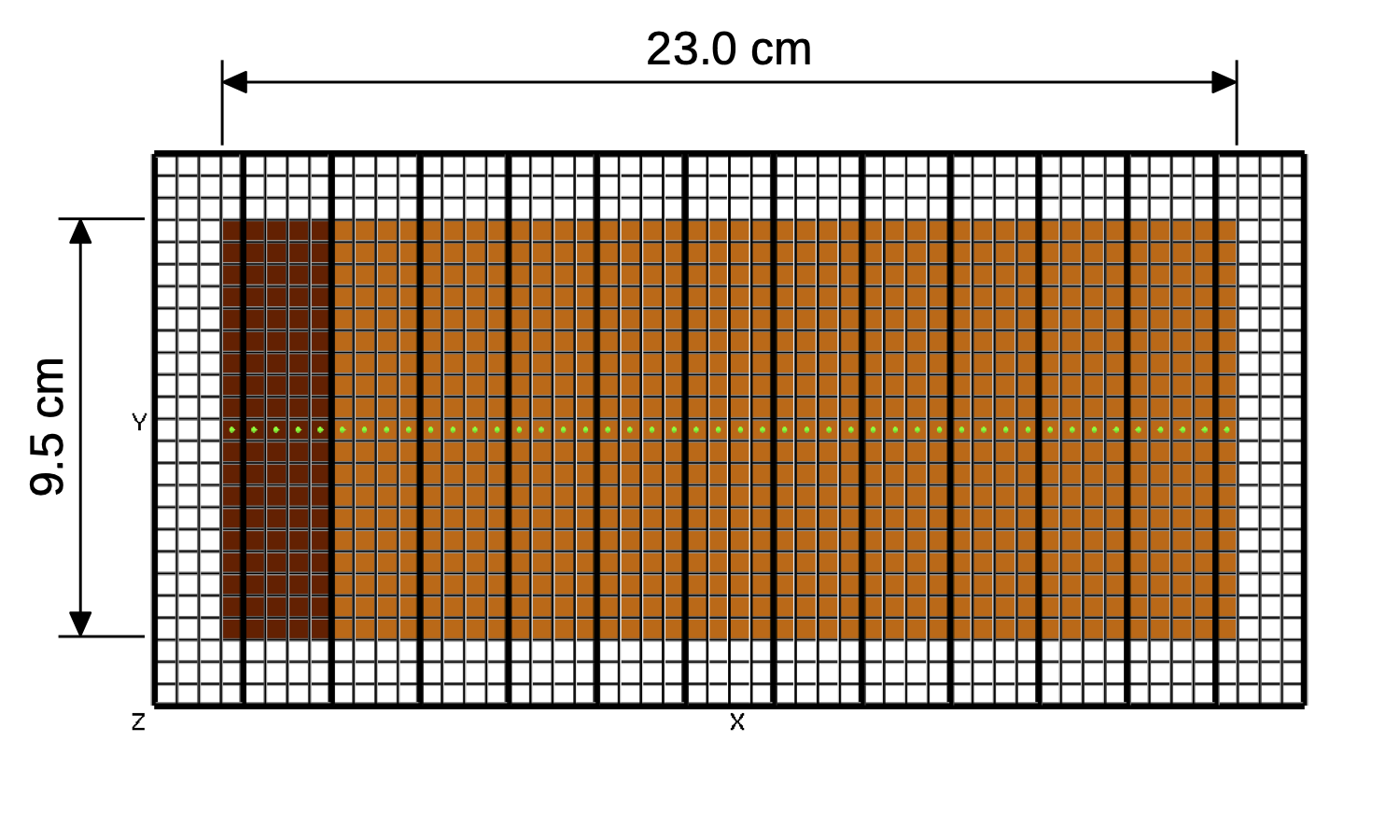}
         \caption{Top view.}
         \label{flamespreadb}
     \end{subfigure}
     \hfill
     \begin{subfigure}[b]{0.49\textwidth}
         \centering
         \includegraphics[width=\textwidth]{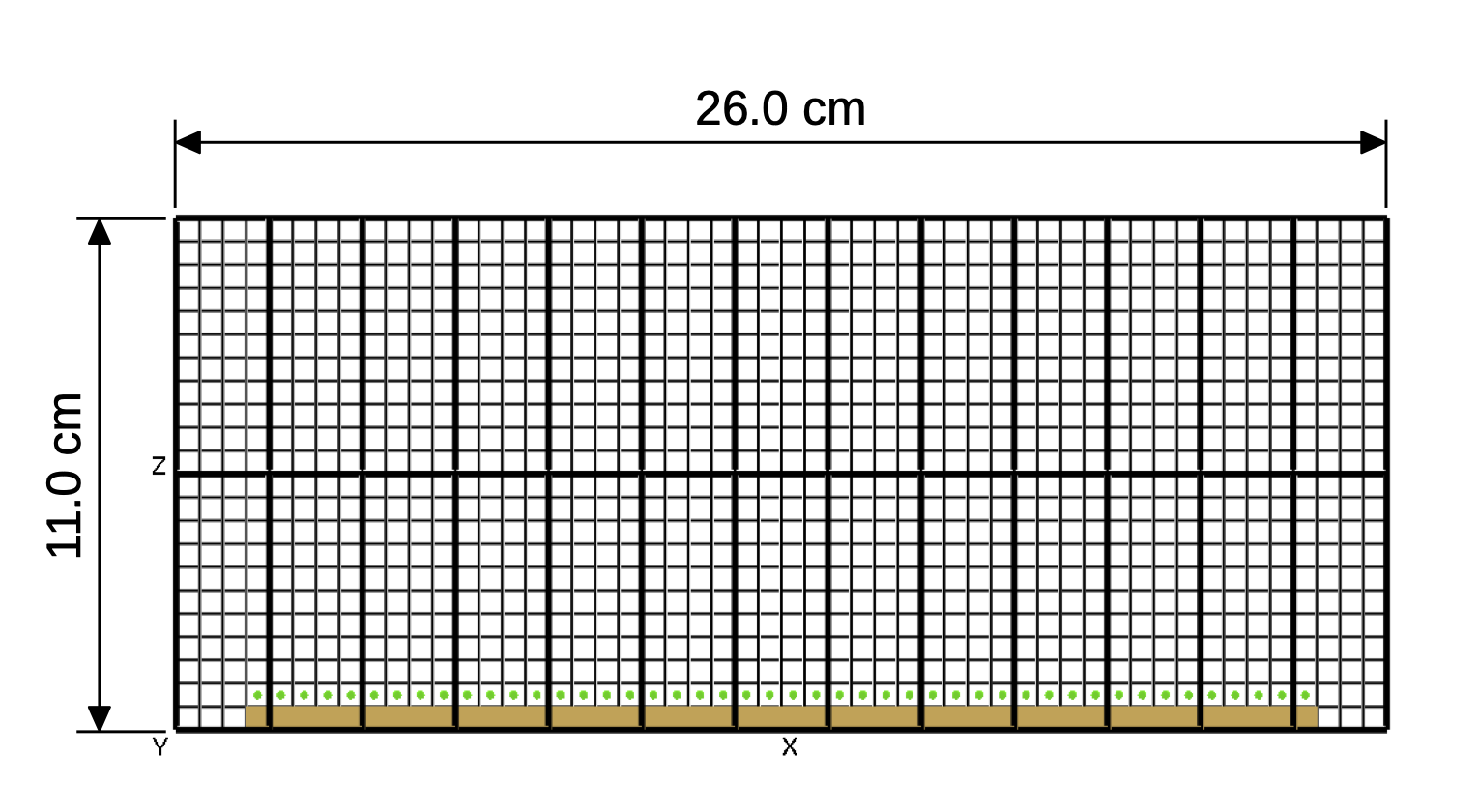}
         \caption{Side view.}
         \label{flamespreadc}
     \end{subfigure}
        \caption{Overview of the horizontal flame spread simulation setup used in the SA, showing the fluid phase resolution, meshes and dimensions. Bold lines indicate mesh borders.}
        \label{flamespread}
\end{figure}

In the flame spread setup, the PMMA sample is positioned on top of an inert ochre obstruction which serves as a sample holder, see Figures~\ref{flamespreada}~and~\ref{flamespreadb}. The sample dimensions are 23~cm~x~9.5~cm~x~0.6~cm, and an external heat flux of 65~kW/m$^2$ is prescribed for 100~s to an area of 2.5~cm~x~9.5~cm to start ignition. The ignition area is located at the left end of the sample, and it is represented by the dark brown patch in Figures~\ref{flamespreada}~and~\ref{flamespreadb}. 

The dimensions of the computational domain are 26~cm~x~12.5~cm~x~11~cm and it is divided in 26~meshes of 2.0~cm~x~12.5~cm~x~5.5~cm each, to allow parallel computation. The fluid cell size used in the flame spread simulation is set to 0.5~cm. The fluid phase modelling differs from the Cone Calorimeter setup only by the cell size. The simulation mode, combustion and radiation modelling are given also according to what was used in the Cone Calorimeter, as described in section~\ref{comb}. An overview of the horizontal flame spread setup is presented in Figure~\ref{flamespread}.

Given its dimensions, the horizontal flame spread simulation can be seen as a simplified small-scale setup of the same scale as the Cone Calorimeter simulation. In addition, the heat flux applied to the dark brown patch is intentionally set to the value used in the Cone Calorimeter, 65~kW/m$^2$. Such similarities allow for the flame in the flame spread simulation to transition from the ignition stage, where the sample heating is modelled according to the Cone Calorimeter, to a self-sustained stage of spread.

\begin{figure}[!htp]
     \centering
     \includegraphics[width=\textwidth]{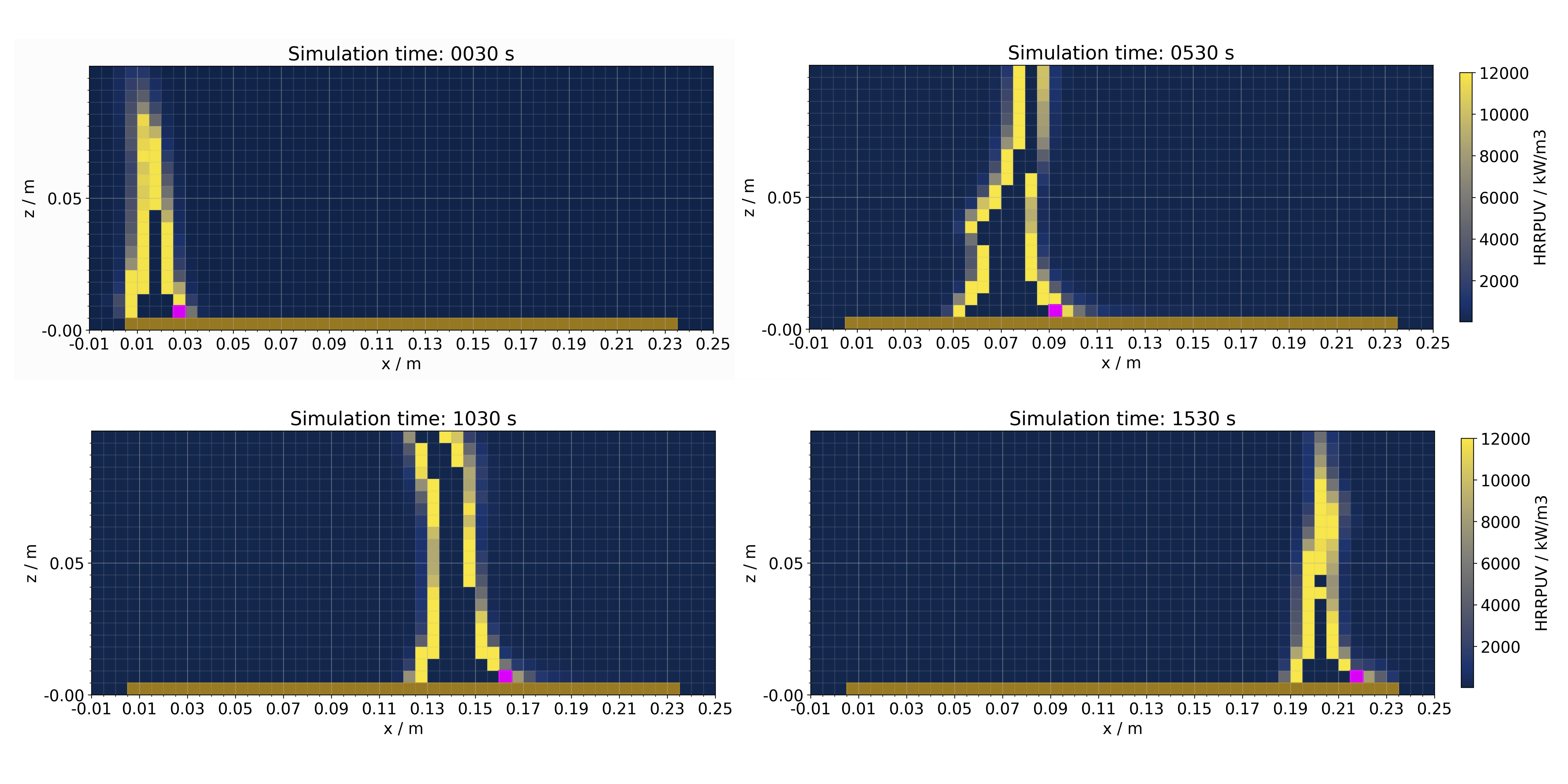}
        \caption{Slices at $y=0$ showing the HRRPUV of the reference case of flame spread simulation, at different points in time. The set of material properties used in the solid phase modelling was transferred from Cone\_04. The fluid cells coloured in magenta indicate the position of the flame front.}
        \label{fsscreenshots}
\end{figure}

In order to build a reference case also for the flame spread simulation, the set of material properties used in Cone\_04 was transferred to the horizontal flame spread setup. The fluid cell size was set to 0.5~cm because it was the minimum tested size to allow a self-sustained spread to occur over the sample. Several other attempts with reasonably larger fluid cell sizes (3~cm, 2~cm, 1~cm) were carried out for even larger samples, but the flame would not spread much farther than the ignition zone, extinguishing shortly after the end of the applied external heat flux. Figure~\ref{fsscreenshots} shows slices containing data of the HRR per unit volume (HRRPUV) taken at the central plane of the domain ($y=0$) along the x~axis, and at different points in time. The patches coloured in magenta in the slices show the location of the flame front, which is here defined as the cell containing the maximum value of HRRPUV in the one-dimensional row of fluid cells touching the surface of the sample. The HRRPUV slices shown in Figure~\ref{fsscreenshots} were generated using the fdsreader version 1.9.9, an open source Python module developed to read FDS output data~\cite{fdsreader}.

\subsubsection{Mean rate of spread (MRS)} \label{mrscalc}

In this section, a methodology is introduced to determine the MRS, since it is not a direct output of the simulation software, and requires an additional post-processing step. The MRS is an important quantity to represent the flame spread phenomenon in the context of fire safety, because it describes how fast a fire can develop in a compartment, impacting the degree of damage and the time to reach flashover. For this reason, the MRS is taken as output of interest in the SA, such that the influence of the material properties can not only be evaluated on the temporal development of the simulated HRR, but also on a single value representing the whole phenomenon. Furthermore, the MRS could be used for validation or as target in future IMPs that aim at estimating input parameters from flame spread experiments in bench-scale dimensions.

\begin{figure}[!htp]
     \centering
     \begin{subfigure}[b]{0.49\textwidth}
         \centering
          \includegraphics[width=\textwidth]{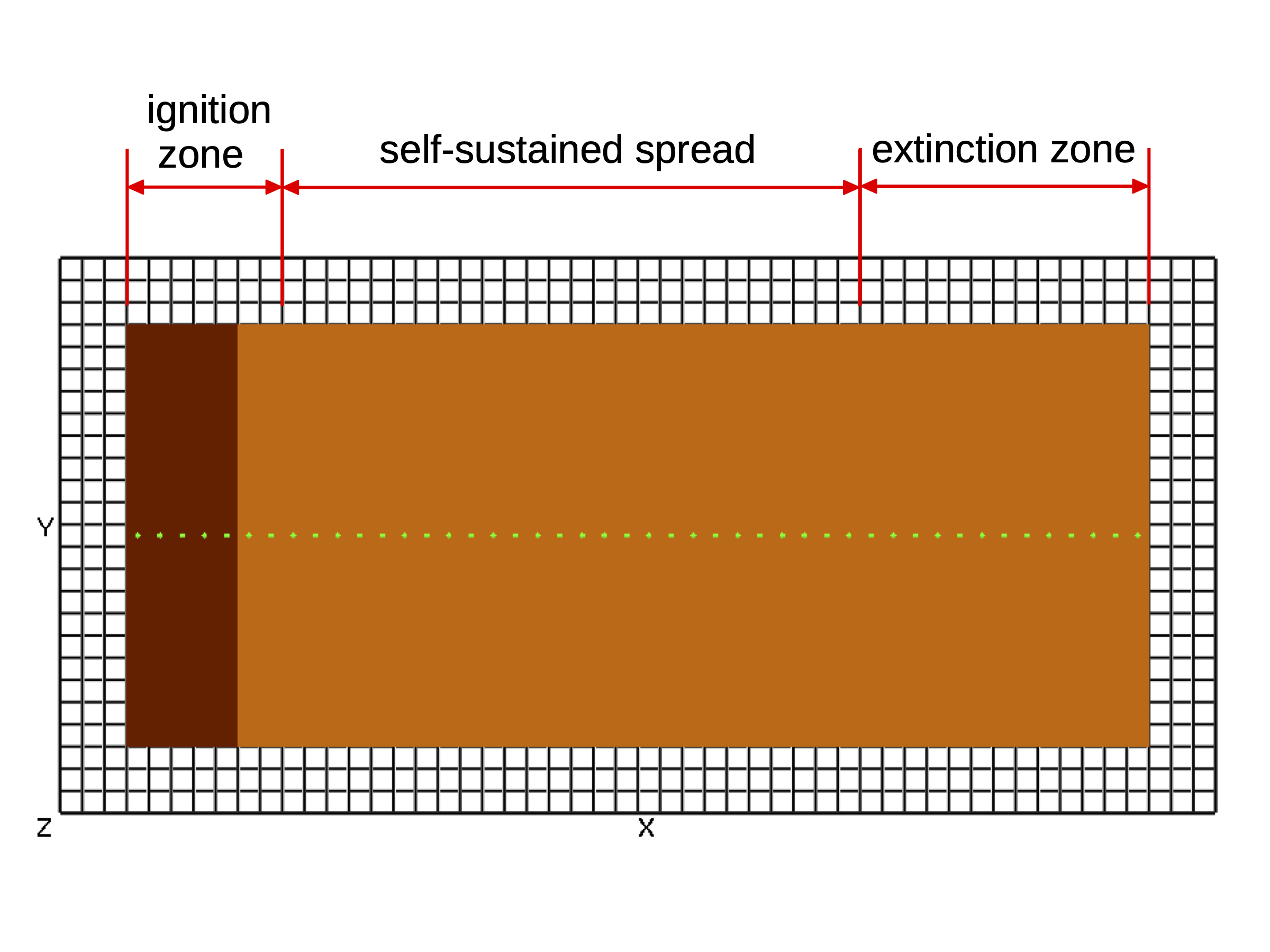}
         \caption{Regimes of spread.}
         \label{mrsa}
     \end{subfigure}
     \hfill
     \begin{subfigure}[b]{0.49\textwidth}
         \centering
         \includegraphics[width=\textwidth]{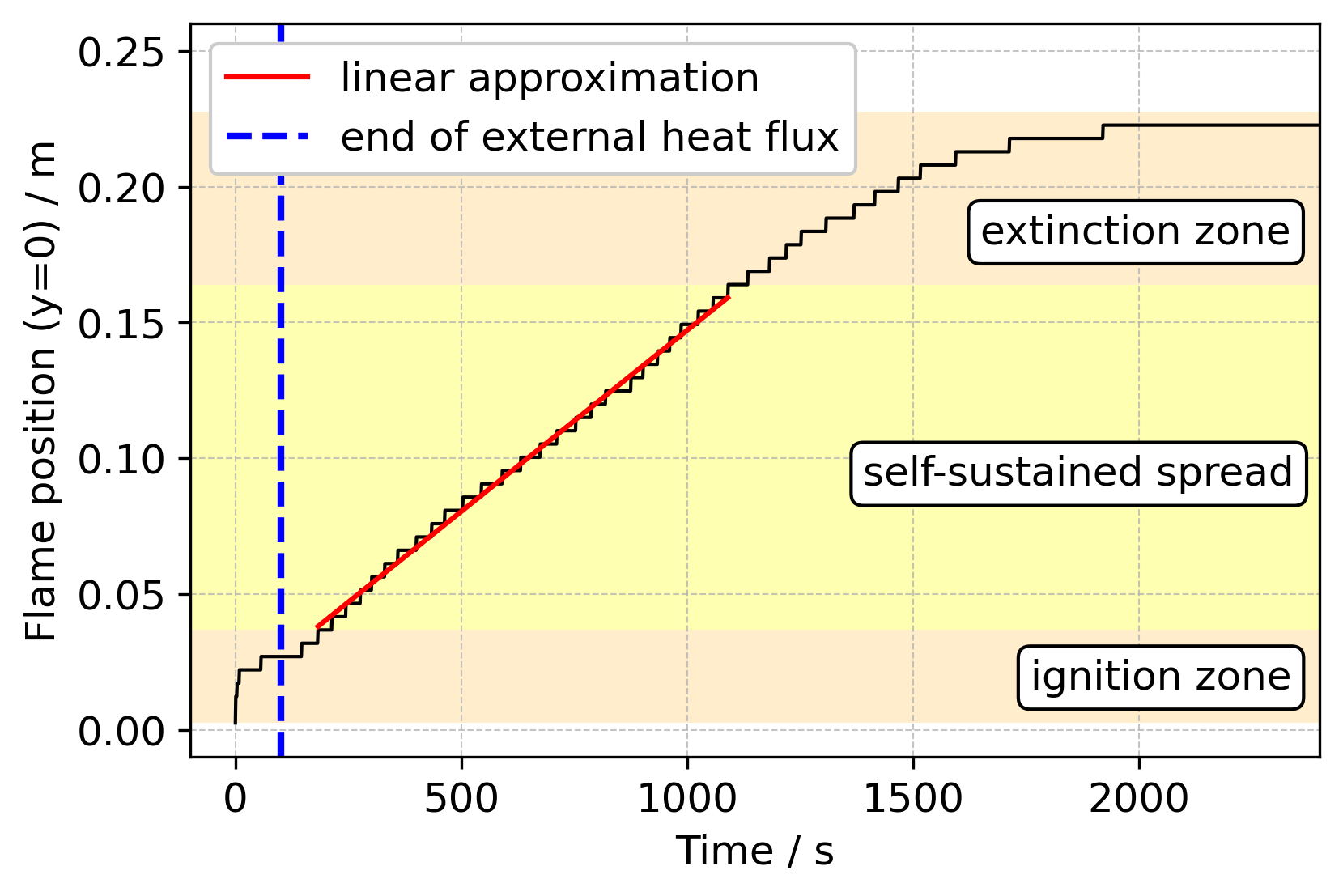}
         \caption{Flame front position \emph{vs} time.}
         \label{mrsb}
     \end{subfigure}
        \caption{Determining the mean rate of spread (MRS) for the reference case of the horizontal flame spread simulation.}
        \label{mrs}
\end{figure}

The MRS is determined by the rate at which the position of the flame front changes with respect to time. At every time step, the maximum HRRPUV is tracked in the fluid cells ahead of the last recorded flame front position, in order to distinguish between leading and trailing edges of the flame. This is done in order to prevent a maximum HRRPUV located in the back of the flame from being mistakenly accounted as the front. In the post-processing, the values of HRRPUV can be read either from slice files, as shown in Figure~\ref{fsscreenshots}, or from multiple devices that are positioned along the centre line of the sample, as indicated by the green dots in Figures~\ref{flamespread}~and~\ref{mrsa}.

The recorded positions can then be plotted against time, as demonstrated in Figure~\ref{mrsb} for the reference case. The small plateaus in the black curve correspond to the periods when the position of the flame front does not change. The vertical blue dashed line at 100~seconds indicates the time instant at which the applied external heat flux over the ignition area ceases. Three zones of spread are identified, as shown in Figures~\ref{mrsa}~and~\ref{mrsb}: 1)~ignition zone; 2)~self-sustained spread; and 3)~extinction zone, influenced by the end of the sample. A linear relation between position and time is fitted to the region where a steady self-sustained spread develops without the influence of the ignition and extinction zones. Finally, the MRS is taken as the slope of the red linear curve in Figure~\ref{mrsb}. In this case, the MRS calculated with the presented method resulted in 0.13~mm/s, which is a value of the same order of magnitude (0.10~mm/s) as those obtained from horizontal flame spread experiments over cast PMMA samples of similar thickness and under the same ventilation conditions, as reported elsewhere~\cite{korobeinichev2018experimental}.

\subsection{Sensitivity Analysis (SA)} \label{sa}


The SAs conducted in this study follows the methodology of the Sobol sensitivity indices, a global SA method based on the decomposition of variances~\cite{sobol2001global}. Unlike local SA methods, which vary input variables one-at-a-time (OAT), the Sobol indices can efficiently explore multi-dimensional parameter spaces, accounting for the effects of all possible combinations of input parameters. Such feature makes it well-suited for quantifying sensitivities of non-linear models, which cannot be adequately assessed by OAT methods~\cite{saltelli2019so}. 

\subsubsection{Mathematical formulation} \label{mathform}


In the mathematical formulation of the method proposed by Sobol~\cite{sobol2001global}, the model can be described as a multivariable function $Y=f(\mathbf{X})$, where~$Y$ is a scalar output and $\mathbf{X}=(X{_1},X{_2},...,X{_k})$ is a point in a $k$-dimensional parameter space. If~$f$ satisfies according requirements, it can be decomposed into terms of increasing dimensions:
\begin{equation}
    f=f_0+\sum_i f_i(X{_i})+\sum_i \sum_{j>i} f_{i j}(X{_i},X_{j})+\ldots+f_{12}\ldots k \quad . 
    \label{eqfdecomp}
\end{equation}
Each individual term is a function only of the inputs in its index, that is, $f_0$ corresponds to the constant part of the function, $f_i = f_i(X{_i})$ depends only on one component (here $X_i$) of the parameter vector $\mathbf{X}$, $f_{ij} = f_{ij}(X{_i},X_{j})$ depends on two components and so on. The total number of terms is equal to~$2^k$, out of each $k$~terms are called first-order function~$f_i$, $f_{ij}$~are second-order functions, and so forth. Equation~\ref{eqfdecomp} can be squared and integrated to generate the decomposition of variances:
\begin{equation}
    V(Y)=\sum_i V_i(X{_i})+\sum_i \sum_{j>i} V_{i j}(X{_i},X_{j})+\ldots+V_{12 \ldots k}
    \label{eqvardecomp}
\end{equation}
where the total variance of the output~$V(Y)$ is split down into $2^k - 1$~different partial variances, each accounting for fractions of the output variance that is induced by the corresponding input, or combinations of inputs. For example, $V_i(X{_i})$ is the induced variance on the output when $X_i$ is varied alone, while $V_{i j}(X{_i},X_{j})$ is the induced variance when $X{_i}$ and $X_{j}$ are varied together. Two parameters are said to interact when their combined effect on the output is different from the sum of their single effects. 

The Sobol indices are obtained by dividing each partial variance in Equation~\ref{eqvardecomp} by the total variance of the output~$V(Y)$, giving:
\begin{equation}
    \sum_i S_i+\sum_i \sum_{j>i} S_{i j}+\ldots+S_{123 \ldots k}=1
    \label{eqind}
\end{equation}
where the indices $S_i$, $S_{ij}$, etc, are ratios varying from 0~to~1. Following a similar terminology used for the decomposition in Equation~\ref{eqfdecomp}, different types of sensitivity indices are defined:
\begin{itemize}
    \item $S_i$: first-order indices, provide a measure of main effects, by varying $X_i$ alone;
    \item $S_{ij..k}$: higher-order indices, measure interaction effects between the inputs indicated in their subscripts;
    \item  $ST_i$: total-order indices, account for all the effects due to variations in $X_i$, i.e. first-order effects and interaction effects.
\end{itemize}
The indices provide a measure of importance of each input parameter by quantifying how much the variance of the output could be reduced if a given input parameter, or combination of parameters, could be fixed. For example, $S_i=0.10$~means that 10\% of the variance of the output could be reduced if $X_i$ is fixed to a known value. Similarly, $ST_i = 0$ implies that $X_i$ is non-influential and can be fixed anywhere in its distribution without affecting the variance of the output~\cite{saltelli2008global}. When interaction effects exist, the sum of all total-order effect indices is greater than unity.

As the number of sensitivity indices to be computed in Equation~\ref{eqind} increases exponentially with the number of input parameters, the calculation of high-order indices can become expensive. It is therefore convenient and often sufficient to express sensitivities in terms of first-order and total-order sensitivity indices. The difference between $ST_i$ and $S_i$ provide a measure of how much $X_i$ is involved in interactions with any other input parameter. 

For a better understanding, an example is given for a function of the form $Y=f(\mathbf{X})$ with $\mathbf{X} = (\textrm{A,B,C})$, for which Equation~\ref{eqind} becomes:
\begin{equation}
     S_\textrm{A} + S_\textrm{B} + S_\textrm{C} + S_\textrm{AB} + S_\textrm{AC} + S_\textrm{BC} + S_\textrm{ABC} = 1
\end{equation}
where $S_\textrm{A}$, $ S_\textrm{B}$, and $S_\textrm{C}$ are the first-order indices, accounting for the main effects of A, B, and C respectively. The second-order indices $S_\textrm{AB}$, $S_\textrm{AC}$ and $S_\textrm{BC}$ account for the interaction effects between the pairs of inputs in their subscripts. Accordingly, $S_\textrm{ABC}$ is the third-order index which accounts for the interaction effects on the output when A, B and C are varied together. In this example, the total-order indices of inputs A, B and C are given as:
\begin{equation}
    ST_\textrm{A} = S_\textrm{A} + S_\textrm{AB} + S_\textrm{AC} + S_\textrm{ABC}
\end{equation}
\begin{equation}
    ST_\textrm{B} = S_\textrm{B} + S_\textrm{AB} + S_\textrm{BC} + S_\textrm{ABC}
\end{equation}
\begin{equation}
    ST_\textrm{C} = S_\textrm{C} + S_\textrm{AC} + S_\textrm{BC} + S_\textrm{ABC}
\end{equation}
in which the total effect of $ST_i$ is the sum of all the terms in Equation~\ref{eqind} where the parameter $X_i$ is considered.

In this study, the sensitivities of the simulation setups to the set of 15 input parameters will be evaluated only in terms of first-order~(S1) and total-order indices~(ST). 


\subsubsection{Estimating the Sobol indices} \label{estimate}

For analytical models, the integrals related to the calculation of variances can be solved also analytically. However, this is not the case for the simulation models investigated in this work. The approach used here for estimating the indices assumes that the simulation model is a ``black box'', and no information on model behaviour is known other than what is perceived through variations in the model's inputs and outputs. The partial variances are calculated by quasi-Monte Carlo estimates and therefore a large number of simulations needs to be conducted, one for each sample of model inputs~\cite{saltelli2002making}. 

Input parameters are assumed to be independent and uniformly distributed within their sampling limits, and samples are generated by employing the Saltelli's sampling scheme~\cite{saltelli2002making}, which is based on the Sobol sequence. The Sobol sequence is a type of low-discrepancy quasi-random sequence that creates an efficient space filling sampling of the high dimensional parameter space. The~SAs, including sampling, estimation of the indices and confidence intervals, are carried out within the SALib Python library (version 1.4.5)~\cite{Iwanaga2022, Herman2017}. 

The sampling limits of the 15 input parameters were defined by taking~15\% of variation around the best parameter set determined in the IMP. A restriction is imposed only for the upper limit of emissivity, such that it would not exceed the value of~0.999.  The sampling limits and units of each input parameter are shown in Table \ref{samplinglimits}. In order to achieve better uniformity, the sampling scheme requires the number of samples~N to be generated as powers of~2, i.e. $\textrm{N}=2^q$, particularly when sampling high-dimensional parameter spaces. For the Cone Calorimeter,~$q=17$, and for the flame spread simulation,~$q=15$. Fewer samples are considered for the flame spread simulation due to the significantly enlarged computing time in comparison to the Cone Calorimeter simulation.



\begin{table}[ht]
    \centering
    \caption{Sampling limits and units of the input parameters considered in the SAs.}
    \begin{tabular}{rllcc}
    \toprule
  & Material   &  Parameter & Sampling limits & Unit \\
    \midrule
1 & PMMA     & Emissivity              & [0.799 ; 0.999]   &  -        \\
2 &          & Absorption coefficient  & [6781.9 ; 9175.5]     &  m$^{-1}$ \\
3 &          & Refractive index        & [2.426 ; 3.281]   &  - \\
4 &          & Conductivity at 150~°C  & [0.322 ; 0.436]   &  W$\cdot$m$^{-1}$$\cdot$K$^{-1}$ \\
5 &          & Conductivity at 480~°C  & [0.021 ; 0.028]   &  W$\cdot$m$^{-1}$$\cdot$K$^{-1}$ \\
6 &          & Conductivity at 800~°C  & [3.687 ; 4.988]   &  W$\cdot$m$^{-1}$$\cdot$K$^{-1}$ \\
7 &          & Specific heat at 150~°C & [0.658 ; 0.890]   &  kJ$\cdot$kg$^{-1}$$\cdot$K$^{-1}$ \\
8 &          & Specific heat at 480~°C & [3.237 ; 4.380]   &  kJ$\cdot$kg$^{-1}$$\cdot$K$^{-1}$ \\
9 &          & Specific heat at 800~°C & [6.183 ; 8.366]   &  kJ$\cdot$kg$^{-1}$$\cdot$K$^{-1}$ \\
10 & Residue & Emissivity              & [0.469 ; 0.635]   &  - \\
11 &         & Conductivity            & [3.833 ; 5.186]   &  W$\cdot$m$^{-1}$$\cdot$K$^{-1}$  \\
12 &         & Specific heat           & [5.009 ; 6.777]   &  kJ$\cdot$kg$^{-1}$$\cdot$K$^{-1}$ \\
13 & Backing & Emissivity              & [0.375 ; 0.507]   &  - \\
14 &         & Conductivity            & [2.047 ; 2.769]   &  W$\cdot$m$^{-1}$$\cdot$K$^{-1}$   \\
15 &         & Specific heat           & [3.457 ; 4.677]   &  kJ$\cdot$kg$^{-1}$$\cdot$K$^{-1}$ \\
    \bottomrule
    \end{tabular}
    \label{samplinglimits}
\end{table}




The influence of the 15 input parameters on two types of outputs are evaluated, namely ``multiple-value'' or ``single-value'' outputs. The simulated HRR is a time-series, therefore it is a multiple-value output. In this case, the sensitivity indices are calculated at every point in time, and, consequently, are presented also as time-series. This is convenient to evaluate how the influence of a given input parameter varies over the course of the simulation.

With respect to the single-value outputs, two indirect quantities are calculated from the simulated HRRs: the root mean square error (RMSE) for the Cone Calorimeter; and the MRS for the flame spread simulation, see section~\ref{mrscalc}.  It is important to evaluate the effect of the input parameters on the RMSE because it is typically taken as the cost function in the optimisation~\cite{lauer2020role}. Here, the RMSE is calculated as:
\begin{equation}
\textrm{RMSE}=\sqrt{\frac{1}{\textrm{N}} \sum_{i=1}^\textrm{N}\left(\textrm{simulated\_HRR}_i-\textrm{measured\_HRR}_i\right)^2}
\end{equation}
where~N is equal to the number of points of the HRR curve, and the measured\_HRR is the experimental HRR curve shown in Figure~\ref{cone04c}, which was used as target in the IMP in earlier work~\cite{hehnen2022PMMA}. Table~\ref{outputs} shows for each setup the number of samples, the output of interest taken in the SAs, their type, and how the Sobol indices are presented.


\begin{table}[ht]
    \centering
    \caption{Summary of the different aspects considered in the SAs based on the Sobol indices.}
    \begin{tabular}{ccccc}
    \toprule
  Simulation setup & Number of simulations & Output of interest & Type of output  &  Sobol indices \\
    \midrule
 Cone Calorimeter    & 131,072 & HRR    & multiple-value  &  time-series     \\
 Cone Calorimeter    & 131,072 & RMSE   & single-value    &  bar plots       \\
 Flame spread        & 32,768  & HRR    & multiple-value  &  time-series     \\
 Flame spread        & 32,768  & MRS    & single-value    &  bar plots       \\
    \bottomrule
    \end{tabular}
    \label{outputs}
\end{table}

The Sobol indices lead to a straightforward and concise way of ranking the input parameters according to their importance, by providing a quantitative measure of sensitivities. This is advantageous when dealing with multi-dimensional parameter spaces. Yet, the indices do not provide the type of relation (i.e. linear, non-linear) between model output and the individual inputs, which is also a meaningful aspect of the analysis. This gap is filled by a complementary qualitative analysis with scatter plots. The scatter plots aim at showing the relation between the output and the two most influential parameters that have been previously identified by the indices. Only single-value outputs (RMSE and MRS) are contemplated in the analysis with scatter plots.

\section{Results and Discussion}\label{sec:results}

\subsection{Effects on the HRRs} \label{resultsHRR}

\begin{figure}[!htp]
     \centering
     \begin{subfigure}[b]{\textwidth}
         \centering
         \includegraphics[width=\textwidth]{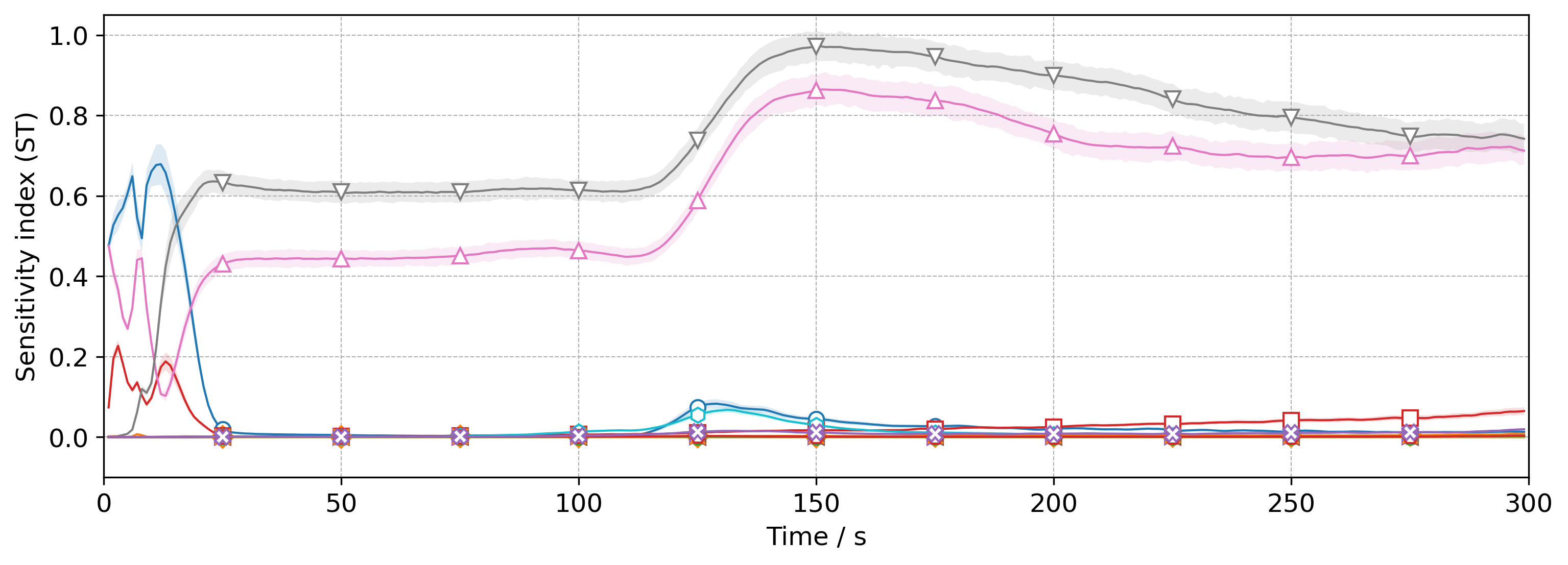}
         \caption{Total-order effects on the HRR of the Cone Calorimeter simulation.}
         \label{stindicesa}
     \end{subfigure}
     \hfill     
     \begin{subfigure}[b]{\textwidth}
         \centering
         \includegraphics[width=\textwidth]{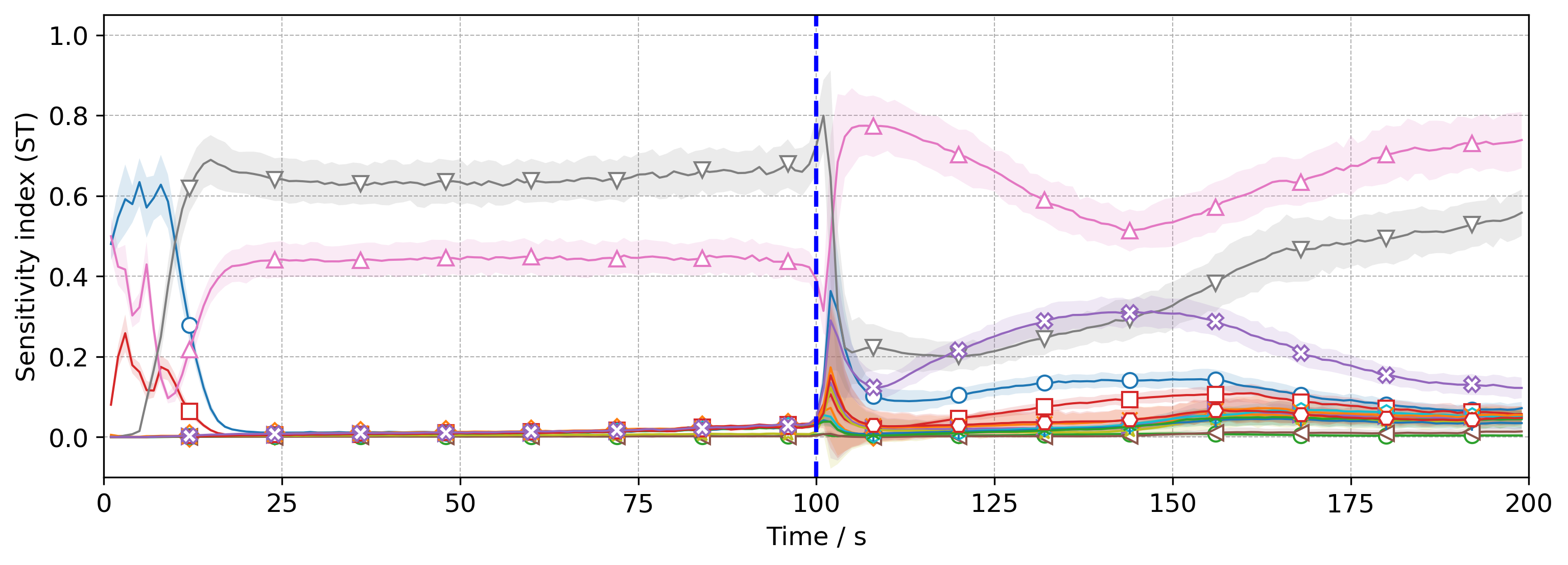}
         \caption{Total-order effects on the HRR of the flame spread simulation (zoom up to 200~seconds).}
         \label{stindicesb}
     \end{subfigure}
     \hfill     
    \begin{subfigure}[b]{\textwidth}
         \centering
         \includegraphics[width=\textwidth]{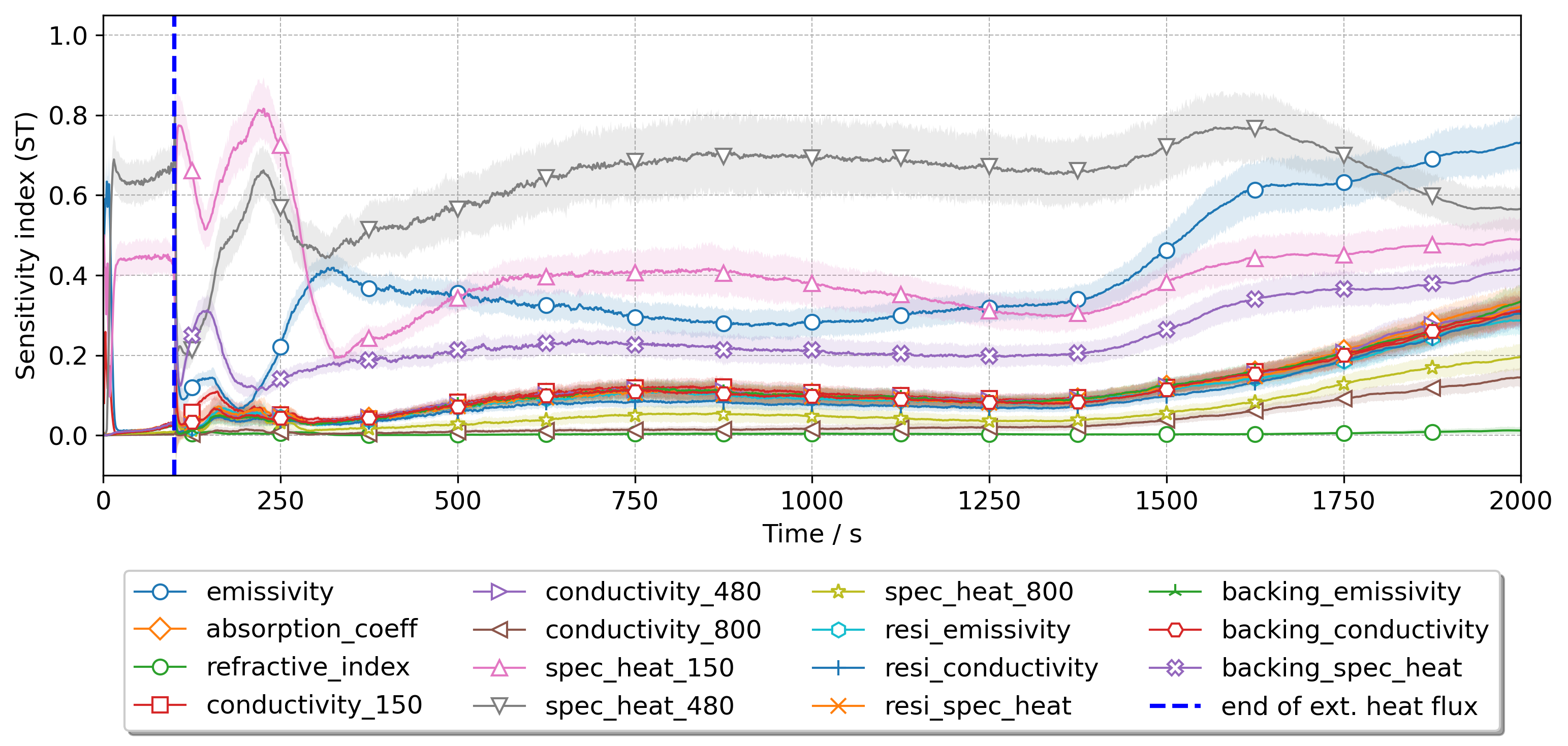}
         \caption{Total-order effects on the HRR of the flame spread simulation.}
         \label{stindicesc}
     \end{subfigure}
        \caption{Time-series of ST indices, indicating total-order effects on the HRRs. The vertical blue dashed line in plots (b) and (c) highlight the end of the external heat flux at 100~seconds in the flame spread simulation.}
        \label{stindices}
\end{figure}

\begin{figure}[!htp]
     \centering
     \begin{subfigure}[b]{\textwidth}
         \centering
         \includegraphics[width=\textwidth]{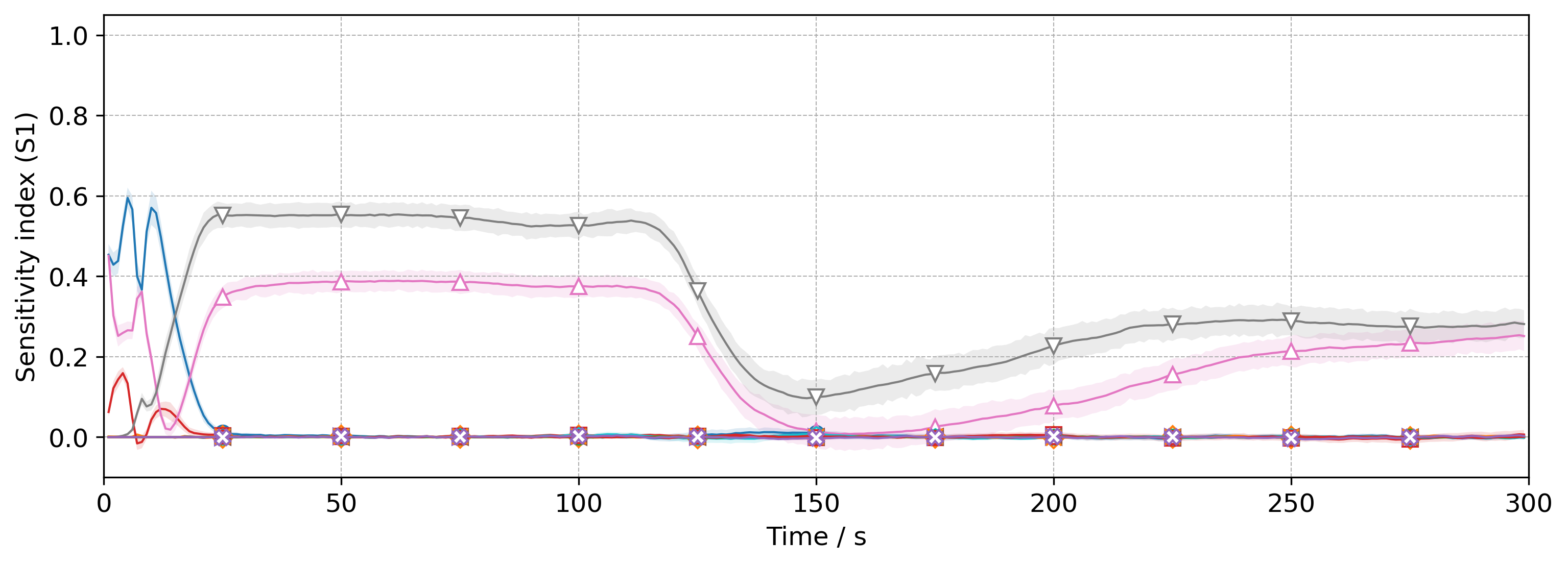}
         \caption{Main effects on the HRR of the Cone Calorimeter simulation.}
         \label{s1indicesa}
     \end{subfigure}
     \hfill     
     \begin{subfigure}[b]{\textwidth}
         \centering
         \includegraphics[width=\textwidth]{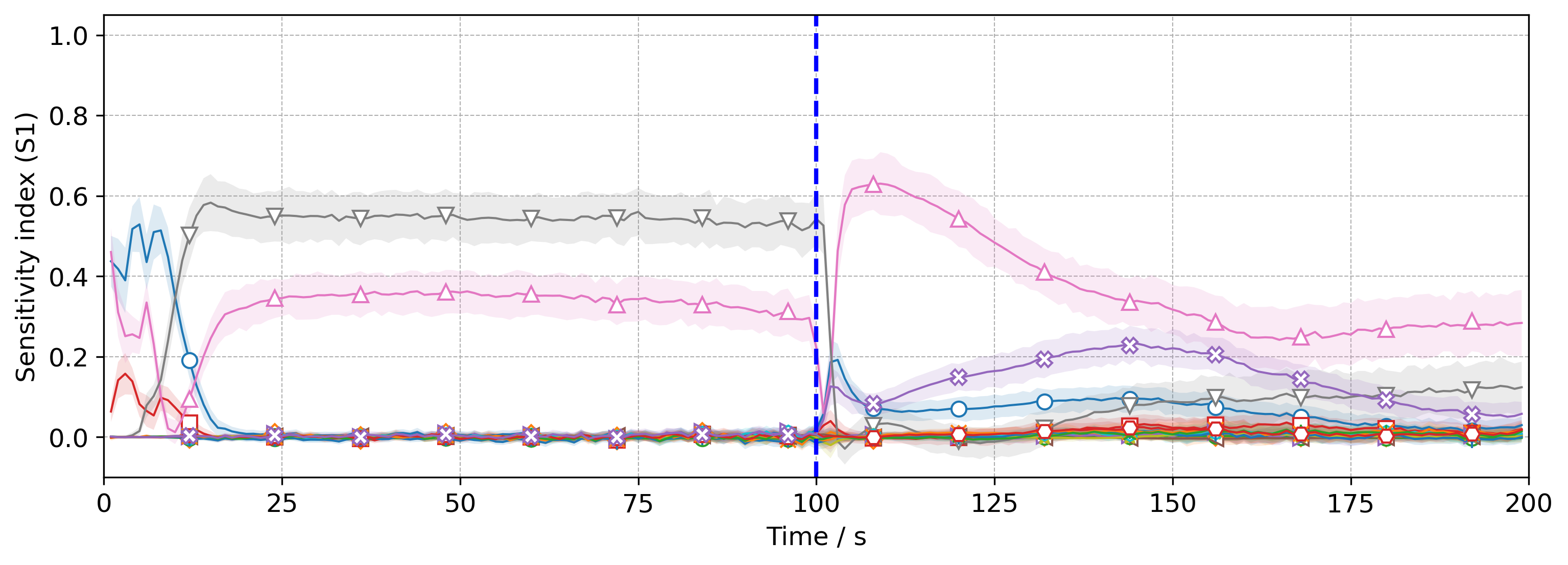}
         \caption{Main effects on the HRR of the flame spread simulation (zoom up to 200~seconds).}
         \label{s1indicesb}
     \end{subfigure}
     \hfill     
    \begin{subfigure}[b]{\textwidth}
         \centering
         \includegraphics[width=\textwidth]{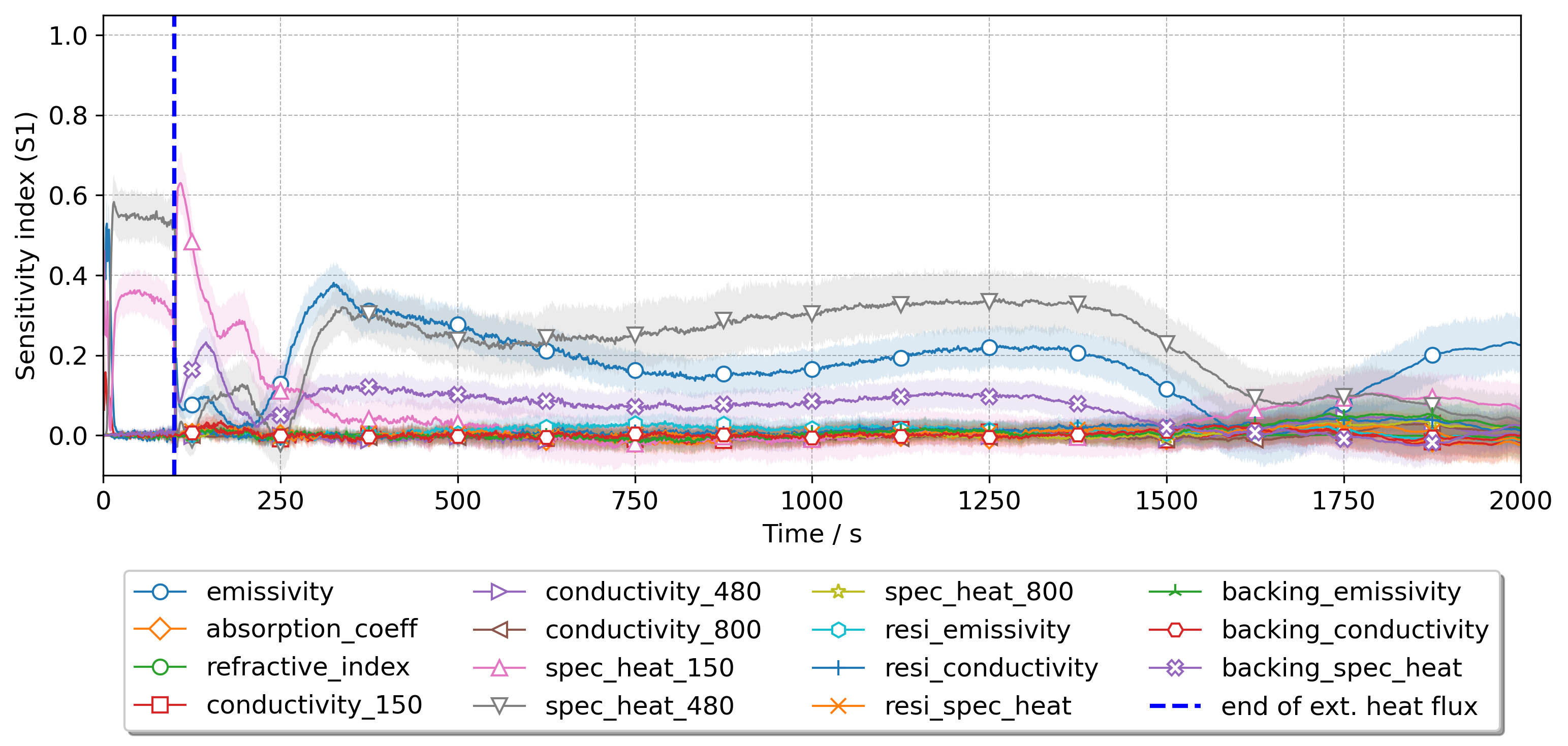}
         \caption{Main effects on the HRR of the flame spread simulation.}
         \label{s1indicesc}
     \end{subfigure}
        \caption{Time-series of S1 indices, indicating main effects on the HRRs. The vertical blue dashed line in plots (b) and (c) highlight the end of the external heat flux at 100~seconds in the flame spread simulation.}
        \label{s1indices}
\end{figure}

The effects of the 15 input parameters on the HRRs of the Cone Calorimeter and the flame spread simulations are discussed in terms of the time-series of the~ST and~S1 indices, expressing respectively the total-order and the main effects of each parameter. The~ST and~S1 indices calculated over the simulated HRR of the Cone Calorimeter simulation
are shown in Figures~\ref{stindicesa}~and~\ref{s1indicesa}. Figures~\ref{stindicesb}~and~\ref{s1indicesb} show the ST and S1 indices with respect to the HRR of the flame spread simulation up to the initial 200~seconds, and Figures~\ref{stindicesc}~and~\ref{s1indicesc} show the indices up to 2000~seconds. 

In Figures~\ref{stindices}~and~\ref{s1indices}, confidence intervals are expressed by the shaded areas around the curves, indicating the uncertainty in the estimation of the indices. The uncertainty in the confidence intervals arises from the inherent variability in the sampled data and the statistical methods employed, reflecting the precision of the Sobol index estimates. Wider intervals indicate greater uncertainty due to limited sample size, while narrower intervals, as observed in the Cone Calorimeter case, result from a larger number of samples, enhancing the reliability of the estimates through the quasi-Monte Carlo approach. 



Regarding the Cone Calorimeter simulation setup, the SA was conducted for the C2, C3 and C5 cases of fluid cell sizes shown in Figure~\ref{conegrid}. As expected, the obtained sensitivity indices were very similar, showing that the HRR response to the input parameters is maintained across the evaluated fluid grid resolutions. The C7 case was left out due to the enlarged computing time that would have been necessary to run the simulations. Nonetheless, due to the similarity in the HRR shapes presented Figure~\ref{conegrid}, it is assumed that the sensitivities to input parameters would be also similar for the C7~case. In light of the similarity and for the sake of brevity, only the SA results with respect to the reference case~(C3) of the Cone Calorimeter will be presented here. The remaining results can be found in the supplementary material data available online~\cite{zenodo:ArticleDataset}.

As can be seen in Figure~\ref{stindicesa}, within the initial 5~seconds of the Cone Calorimeter simulation, only three parameters have non-zero ST~indices: PMMA emissivity, conductivity at 150~°C and specific heat at 150~°C.  This means that these are effectively the only parameters affecting the HRR up to this point in time, whereas all the other parameters remain unimportant. At about 10~seconds, the importance of the specific heat at 150~°C momentarily drops, while the influence of the same property at 480~°C increases. This is explained by the dependency of the specific heat on temperature, established by piecewise linear function. As the sample heats up, higher temperatures are reached, causing the value of the property to change, and consequently its influence over the HRR. Soon after that, at about 25~seconds, the importance of the specific heat at 150~°C increases again, and it becomes, together with the specific heat at 480~°C, the two most important parameters to affect the HRR of the Cone Calorimeter. At the same time, the initial total-order effects of PMMA emissivity and conductivity decrease to a practically negligible value for the rest of the simulation time. 

It is important to distinguish, however, two stages of influence of the values of specific heat on the HRR, which become evident when Figures~\ref{stindicesa} and~\ref{s1indicesa} are compared. The stages are defined by the difference between their respective ST (Figure~\ref{stindicesa}) and S1 (Figure~\ref{s1indicesa}) indices, indicating the degree of interaction effects between the two parameters. In the first stage, ranging from about 25~to~120~seconds, the difference between ST and S1 is small, indicating therefore that the interaction effects are also small. This implies that the HRR is affected predominantly by main effects of the two parameters, expressing that the effects of changing them individually is dominant to the HRR at this stage. In the second stage, from 120~to~150~seconds, an approximately synchronised and significant increase of the ST indices is observed. At the same time, their corresponding S1 indices decline in a similar trend, as presented in Figure~\ref{s1indicesa}. The large discrepancy between~ST and~S1, together with the fact that all the other parameters have ST indices close to zero, indicate that strong interaction effects between the values of specific heat at 150~°C and at 480~°C dominate the HRR from this point forward in the Cone Calorimeter simulation. Still, at around 125~seconds, a timid increase in the ST indices of the PMMA and the residue emissivities are almost solely related to interaction effects, given that their correspondent S1 indices are very close to zero. 



By comparing Figures~\ref{stindicesb}~and~\ref{s1indicesb} to Figures~\ref{stindicesa}~and~\ref{s1indicesa}, a very similar ranking of effects can be identified between the initial 100~seconds of the flame spread simulation, and the initial 125~seconds of the Cone Calorimeter simulation. In the flame spread setup, the initial 100~seconds corresponds to the heating of the dark brown part of the sample by the prescribed external heat flux to start ignition (see Section~\ref{fssetup}). Given that this approach is the same as the one used in the Cone Calorimeter for heating up the whole sample, the similarity in modelling for using the \verb|EXTERNAL_FLUX| function in FDS is clearly captured by the sensitivity indices.

The vertical blue dashed line at 100~seconds in Figure~\ref{stindicesb} marks the end of the external heat flux and thus the transition from the ignition phase to the phase where the spread is self-sustained. The transition from one phase to another is highlighted by an abrupt change in the importance of some parameters. They are the specific heat at 150~°C and at 480~°C, emissivity, conductivity at 150~°C of PMMA, and the specific heat of the insulation material. From Figure~\ref{stindicesc} it can be seen that despite oscillations, after the initial 200~seconds those parameters remain as the most influential ones affecting the HRR. Another important observation concerns the least important parameters, whose effects on the HRR, although smaller, are not insignificant. The only exception is the refractive index, whose indices turned out to be zero in both simulation setups.

The existence of a transition in parameter importance, defined by the interruption of the external heat flux and the start of a self-sustained spread, indicates the differences between the Cone Calorimeter and the flame spread heating conditions. In the Cone Calorimeter, the sample is subject to a considerably higher heat flux than that of the flame spread, where only a small fraction of the heat transferred by radiation reaches the sample. Despite varying in intensity, the HRR of the flame spread is consistently influenced by multiple parameters throughout the simulation (see Figure~\ref{stindicesc}). As it seems not be the case for the Cone Calorimeter simulation, this implies that not all parameters that are important to the flame spread can be well estimated during IMPs based solely on the Cone Calorimeter setup.

\subsection{Effects on the RMSE and MRS} \label{rmse_mrs}

\begin{figure}[!htp]
     \centering
     \begin{subfigure}[b]{0.49\textwidth}
         \centering
          \includegraphics[width=\textwidth]{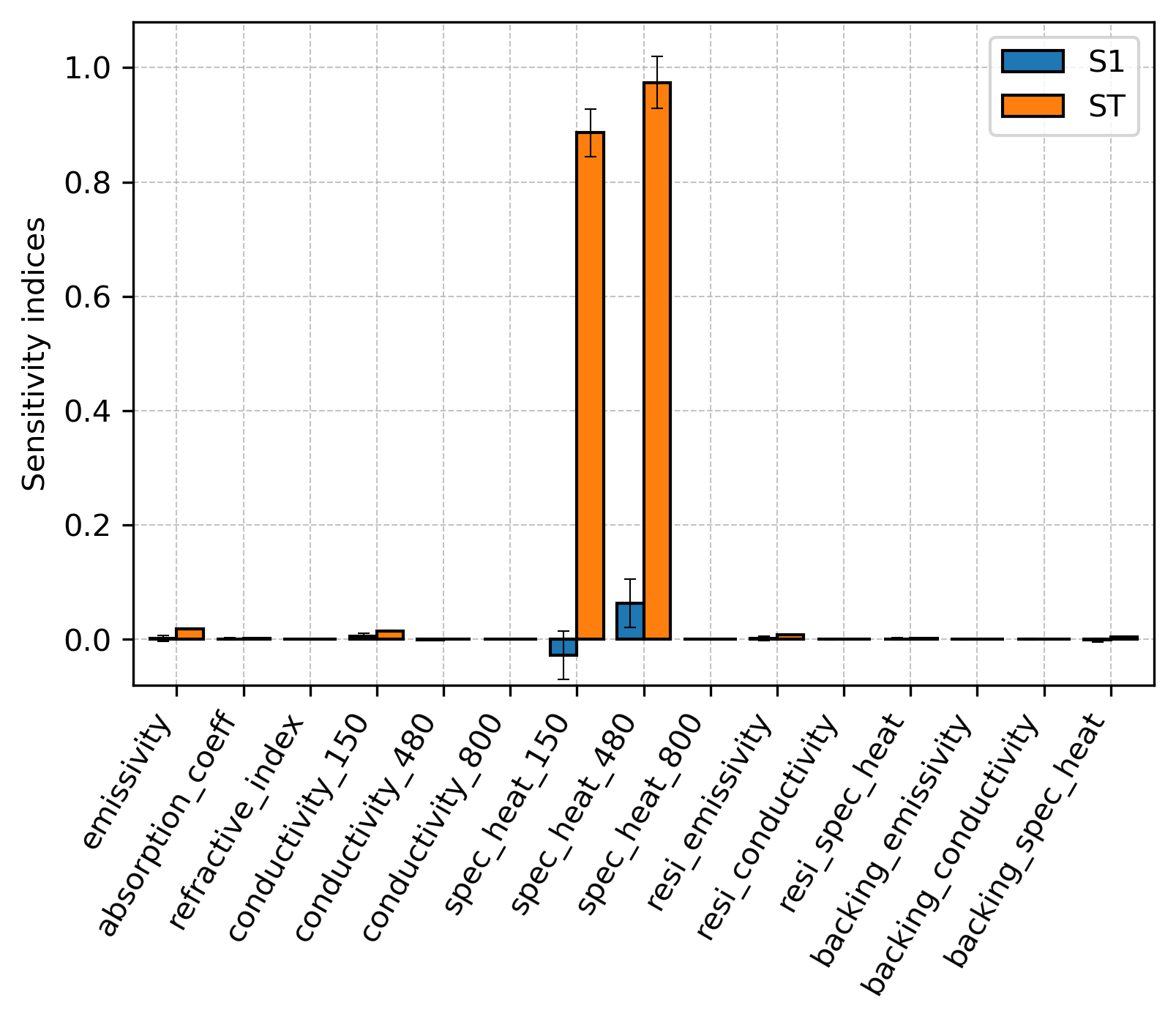}
         \caption{Effects on the RMSE - Cone Calorimeter simulation.}
         \label{sts1rmse}
     \end{subfigure}
     \hfill
     \begin{subfigure}[b]{0.49\textwidth}
         \centering
         \includegraphics[width=\linewidth]{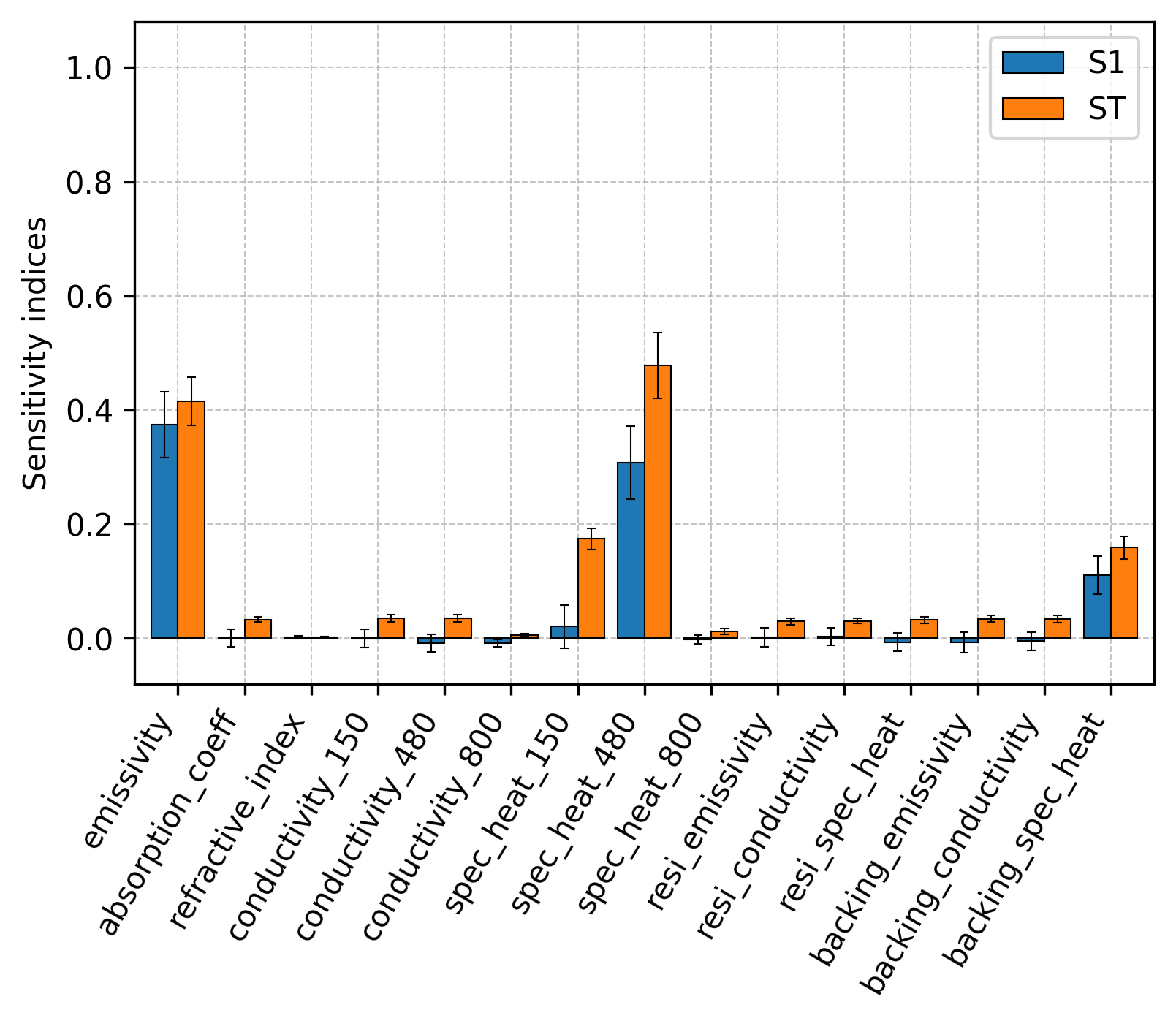}
         \caption{ Effects on the MRS - horizontal flame spread simulation.}
         \label{sts1mrs}
     \end{subfigure}
     \caption{Sobol sensitivity indices indicating the effects of the 15 input parameters on the single-value outputs of the two simulations setups.}
     \label{sts1rmsemrs}
\end{figure}

The effects of the 15~input parameters on the RMSE and on the MRS are presented respectively in Figures~\ref{sts1rmse} and~\ref{sts1mrs}. Since both are single-value outputs, sensitivities are described by a single set of indices for each case. Confidence intervals are indicated by error bars. Some S1~indices present negative values, which are due to numerical artefacts in the estimates. This issue has been reported before as a common characteristic of Saltelli's method that is often associated with the value of the index being close to zero. 

As can be seen from Figure~\ref{sts1rmse}, the specific heat at 150~°C and at 480~°C are effectively the only two input parameters that significantly affect the RMSE (calculated between simulated HRR curves of the Cone Calorimeter and experiment, see Section~\ref{estimate}). Moreover, the influence is characterised by strong interaction effects, due to the negligible~S1 and dominant~ST values. This means that the effect on the RMSE highly depends on how the values of these two parameters are combined. Indeed, the combination between the specific heat at 150~°C and at 480~°C defines the slope of the linear curve that relates the two values in the ramp function. The slope is in turn related to how fast the change in the specific heat will occur as a function of temperature. For example, if the specific heat changes from a low value to a high value abruptly, suddenly more energy is required to cause the temperature of the material to change. In this case, this would reduce the local rates of pyrolysis, slowing down the production of combustible gases which then burn, releasing heat. Ultimately, this translates into flattened HRR curves. This change in the HRR shape determines how much each simulated curve deviates from the experimental HRR curve. This effect is then captured by the RMSE. 

\begin{figure}[!htp]
     \centering
    \begin{subfigure}[b]{0.49\textwidth}
         \centering
          \includegraphics[width=\textwidth]{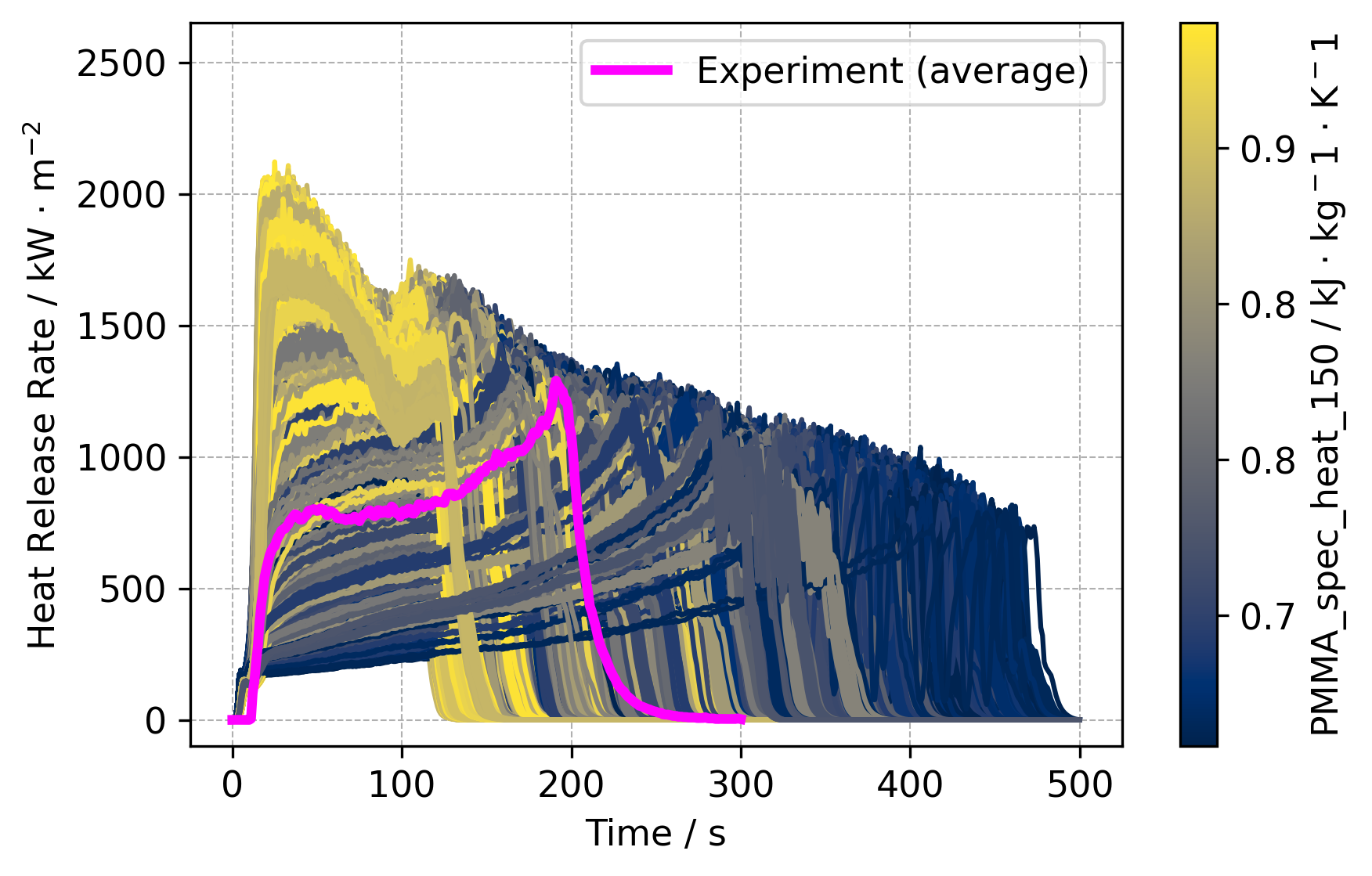}
         \caption{Colour map: samples of specific heat at 150~°C.}
         \label{hrrcurvesconea}
     \end{subfigure}
     \hfill
     \begin{subfigure}[b]{0.49\textwidth}
         \centering
         \includegraphics[width=\linewidth]{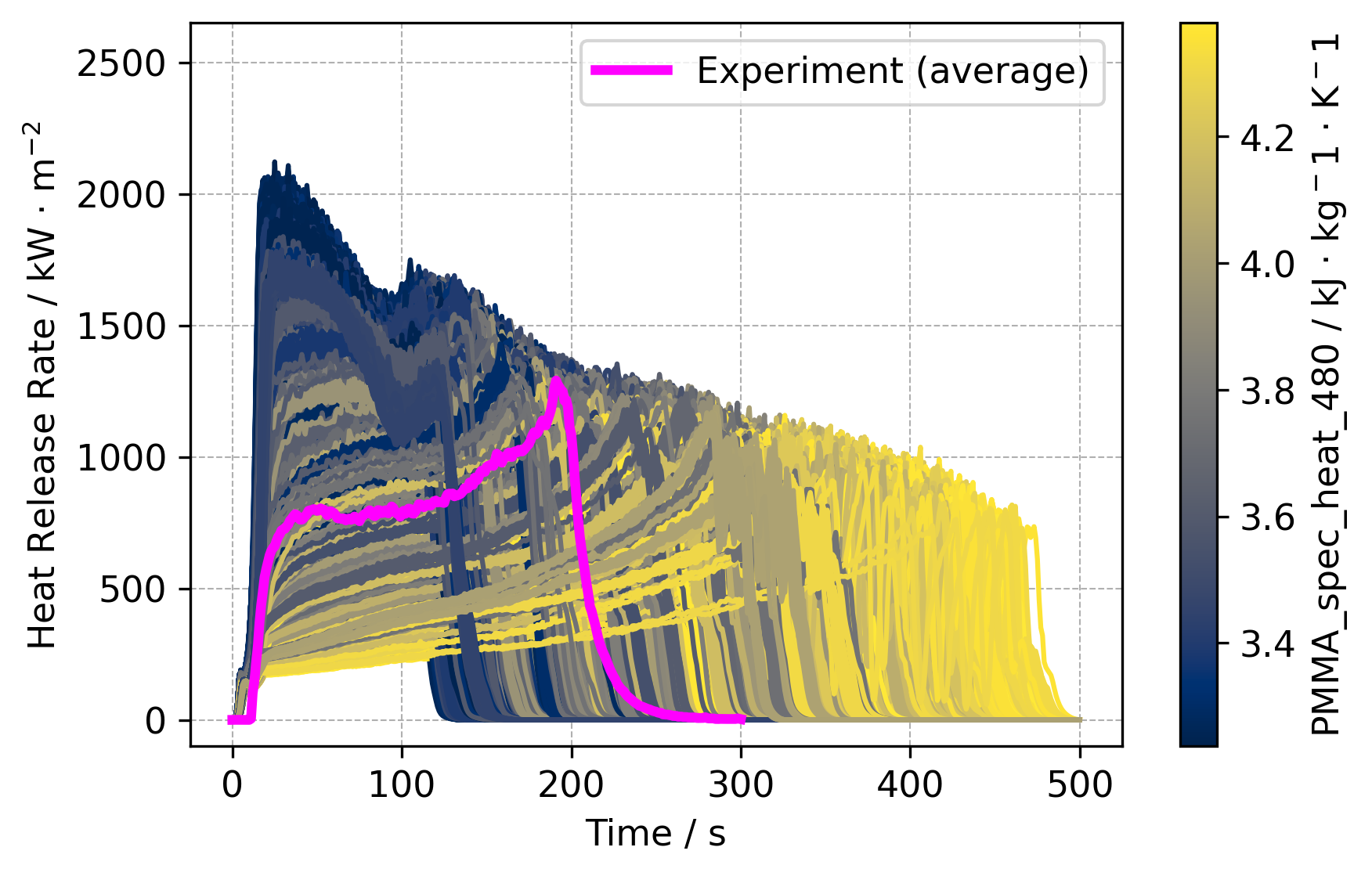}
         \caption{Colour map: samples of specific heat at 480~°C.}
         \label{hrrcurvesconeb}
     \end{subfigure}
        \caption{Visualisation of the interaction effects between the time-dependent values of specific heat  on the HRR curves of the Cone Calorimeter simulation.}
        \label{hrrcurvescone}
\end{figure}

A complementary way to visualise this strong interaction effects is presented in Figure~\ref{hrrcurvescone}. The plots show all the simulated 131,072 HRR curves from the Cone Calorimeter coloured by the corresponding samples of specific heat at 150~°C in Figure~\ref{hrrcurvesconea}, and by the samples of specific heat at 480~°C in Figure~\ref{hrrcurvesconeb} used in each simulation. The colour map illustrates how the shape of the HRR is affected by combinations between values of these two most important parameters. The red curve is the HRR from Cone Calorimeter experiments, taken as reference for determining the RMSE.

In comparison to what is observed for the RMSE, a different scenario of sensitivities is identified for the MRS, as depicted in Figure~\ref{sts1mrs}. The~ST and S1~indices show that PMMA emissivity and specific heat at 480~°C are, in order, the two most important parameters to affect the MRS of the ones investigated here, and no meaningful interaction effects are observed. The higher importance of PMMA emissivity to the MRS is reasonable, given that a horizontal flame spread is controlled mainly by the heat that is radiated from the flame. It would be expected, in situations where the spread is controlled mainly by convection, e.g. vertical spreads, that emissivity would not play the leading role in influencing the MRS. Amongst the input parameters with lower importance, the specific heat of the insulation material and the specific heat of PMMA at 150~°C are the two most important ones. 


At this point, it is important to emphasise what is the main implication to the IMP revealed by the sensitivity indices shown in Figure~\ref{hrrcurvescone}. The RMSE is a common approach to measure deviations between two sets of data, and therefore it is commonly used as cost function during the inverse modelling~\cite{lauer2020role}. This means that whatever influence an input parameter has on the model output (in this case, the HRR of the Cone Calorimeter), it should be reflected in the RMSE for an effective estimation. However, comparison of Figures~\ref{stindicesa} and~\ref{hrrcurvesconea} suggests that the initial importance of PMMA emissivity and specific heat at 150~°C to the HRR (Figure~\ref{stindicesa}) is not manifested to the same level in the RMSE. This implies that not only is it necessary that the direct model output is sufficiently sensitive to the inputs that are important to the flame spread, but also that the cost function is. In this regard, neither of these two requirements were met, since there are several other input parameters affecting the flame spread (reflected both in the HRR and MRS) that have little or no importance to the HRR in the Cone Calorimeter and/ or to the RMSE. 

\begin{figure}[!htp]
     \centering
    \begin{subfigure}[b]{0.49\textwidth}
         \centering
          \includegraphics[width=\textwidth]{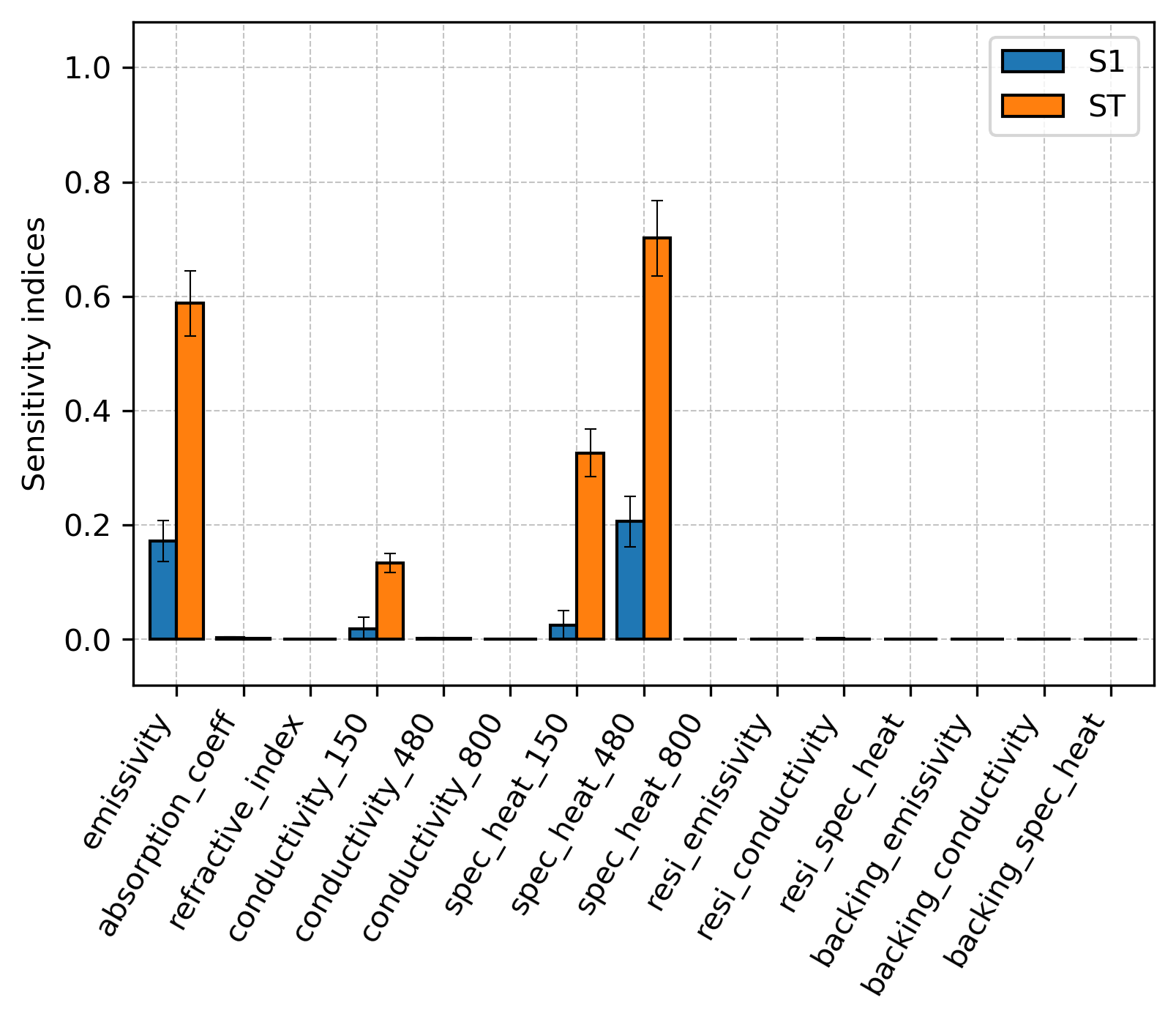}
         \caption{Effects on the RMSE over 0 - 20 seconds.}
         \label{sts1rmsepartsa}
     \end{subfigure}
     \hfill
     \begin{subfigure}[b]{0.49\textwidth}
         \centering
         \includegraphics[width=\linewidth]{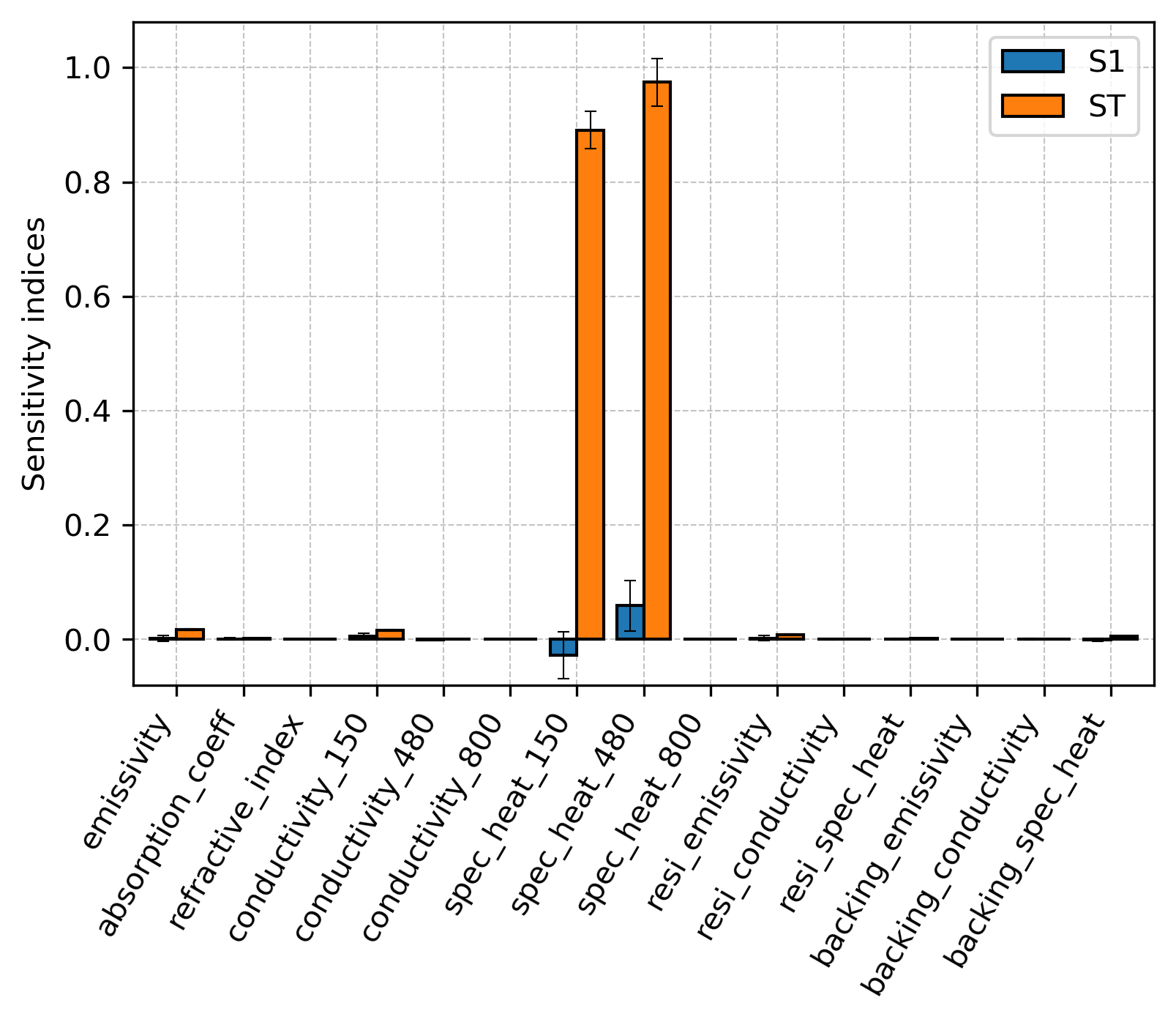}
         \caption{Effects on the RMSE over 20 - 300 seconds.}
         \label{sts1rmsepartsb}
     \end{subfigure}
        \caption{Sobol sensitivity indices indicating the effect of input parameters on the RMSE calculated over different stages of the Cone Calorimeter simulation.}
        \label{sts1rmseparts}
\end{figure}

Motivated by the results shown in Figure~\ref{stindicesa}, where differences exist between the initial 20~seconds and the rest of the simulation, additional SAs were conducted taking two different extracts of the RMSE as output of interest. In the first extract, the RMSE is calculated up to 20~seconds of simulation time (RMSE-0-20) and deviations to the experimental data are calculated accordingly up to the 20th second. Similarly, in the second extract, the RMSE is calculated from 21 to 300~seconds (RMSE-20-300). Sensitivity indices are presented for these two approaches in Figures~\ref{sts1rmsepartsa} and~\ref{sts1rmsepartsb}, respectively. As expected, the results for each extract present a clear correspondence to the different sensitivity profiles shown in Figure~\ref{stindicesa}. The rank of parameter importance up to 20~seconds to the HRR is equivalent to the rank shown in Figure~\ref{sts1rmsepartsa} to the RMSE-0-20. The same is true for the second extract. The dominance of the specific heat values after the 20th second is reflected in the RMSE-20-300. This analysis helps to understand why the default RMSE seems not to be significantly influenced by any other parameter than the specific heat at 150~°C and at 480~°C. Since the importance of these two values is higher for the most part of the simulation, the brief importance of the other parameters (emissivity, conductivity at 150~°C) gets diluted when the whole HRR time-series is condensed in a single RMSE value. 

These observations highlight the importance of making use of cost functions that are as sensitive as the model outputs that they intend to represent. This way, a new design of cost functions can be defined. Instead of optimising for the global RMSE, which may be dominated by only a subset of sensitive parameters, a combination of RMSE at different phases of the experiment may cover the full set of sensitive parameters.
Another possible alternative would be if the MRS could be taken as the cost function instead of the RMSE in IMPs based on bench-scale flame spread experiments. In this case, an approach as the one introduced in Section~\ref{fssetup} could be adopted as the single-value model output to be optimised against a single measured MRS. Yet, this strategy may be computationally expensive and further research on this idea is needed, particularly because it does not exist to date a typical bench-scale flame spread experiment as the Cone Calorimeter.

All in all, in addition to revealing differences between the Cone Calorimeter and the horizontal flame spread setups, the SA on the Cone Calorimeter disclosed that only 4 out of 15 input parameters have non-negligible influences. In terms of thermophysical properties, these four parameters are in fact three: emissivity, conductivity, and specific heat of the PMMA sample. This observation is similar to the results found in the work of Fleurotte et al.~\cite{fleurotte2022sensitivity}, who conducted a SA based on the Morris method to determine which parameters are more important to the HRR of a Cone Calorimeter model. In their work, PMMA emissivity and specific heat capacity are among the most influential parameters, along with activation energy and density. Thus, ranking the parameters according to their importance allows for model simplification, by excluding the non-influential inputs and/ or fixing them in the optimisation. This approach can potentially reduce the computing time by orders of magnitude, depending on the characteristic of the optimisation method used~\cite{hehnen2022PMMA}. 

\subsection{Scatter plots}

\begin{figure}[!htp]
     \centering
     \begin{subfigure}[b]{0.49\textwidth}
         \centering
          \includegraphics[width=\textwidth]{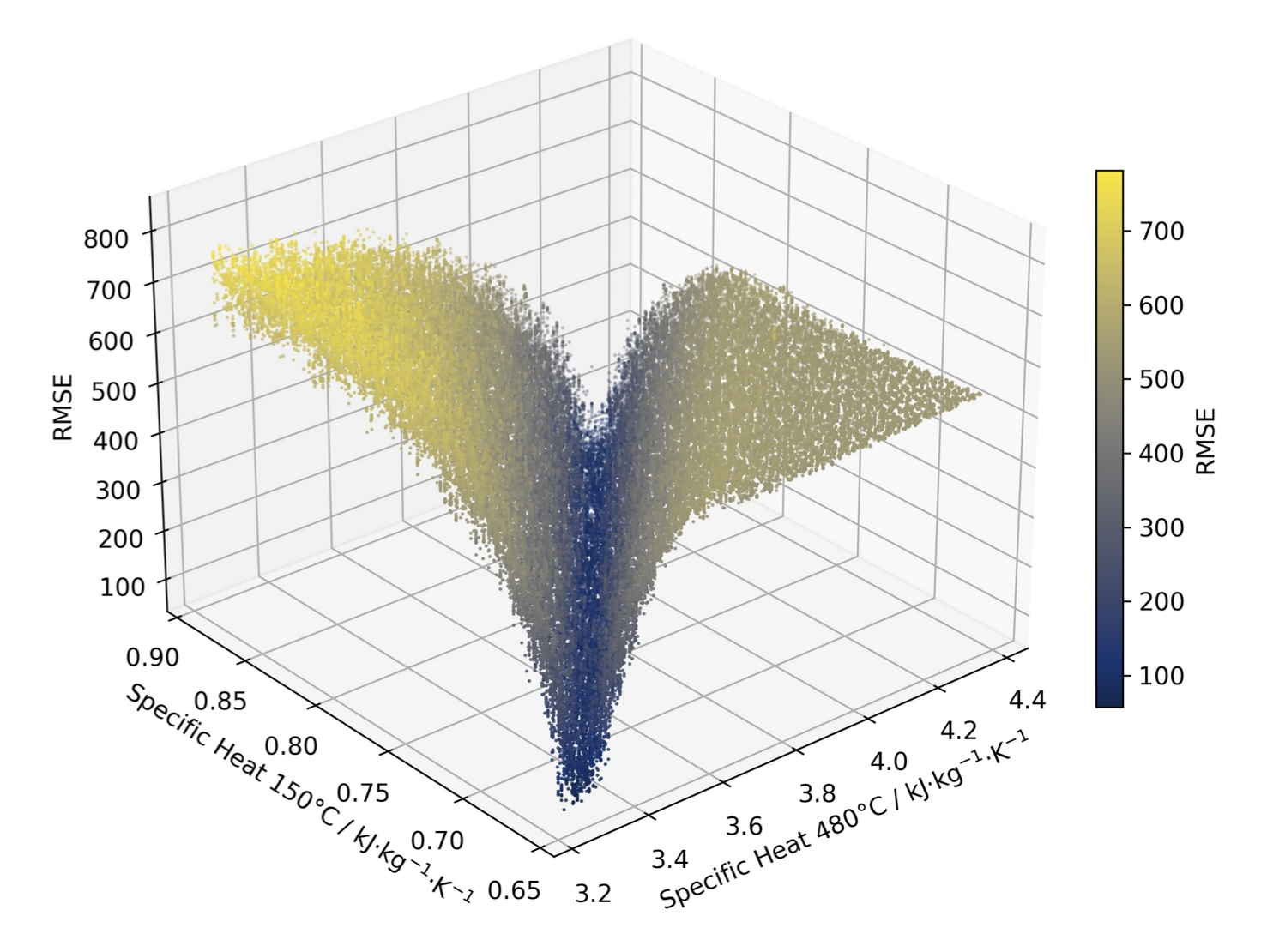}
         \caption{3-D surface.}
         \label{ccscattera}
     \end{subfigure}
     \hfill
     \begin{subfigure}[b]{0.49\textwidth}
         \centering
         \includegraphics[width=\textwidth]{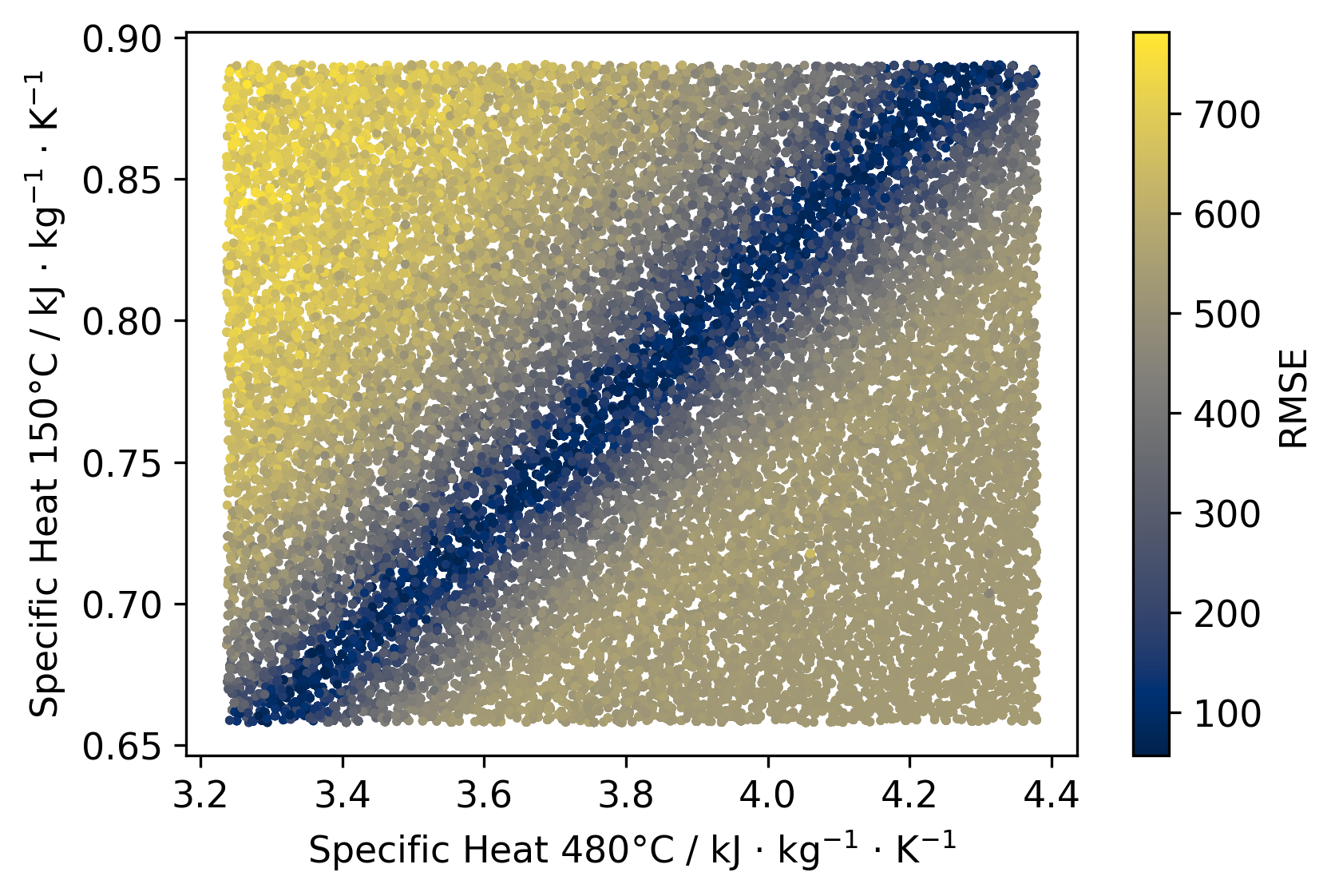}
         \caption{2-D projection.}
         \label{ccscatterb}
     \end{subfigure}
        \caption{Values of RMSE plotted against its two most influential parameters.  }
        \label{ccscatter}
\end{figure}

The calculated values of RMSE in the Cone Calorimeter setup are plotted against the sampled values of the two most influential parameters: the specific heat at 150~°C and at 480~°C. The result is a 3-D plot, where a well-defined surface allows the graphical interpretation of the interaction effects between the two inputs on the RMSE, see Figure~\ref{ccscattera}. The 3-D~surface reveals a dark-blue valley for which the RMSE values are minimised when certain combinations of the two parameters are taken. A 2-D projection over the axes of input parameters shown in Figure~\ref{ccscatterb} clarifies that such combinations belong to a linear shaped subset of samples in their parameter space. Another interesting region in the 3-D~surface is the light green plateau formed by nearly constant values of RMSE. The existence of a plateau reveals a significant portion of the input space that leads to no meaningful change in the RMSE. This is particularly important for the optimisation, because it can decrease its efficiency and lead to convergence to local minima. 

A very similar relation to the one shown in Figure~\ref{ccscatter}, is presented in the work of \cite{batiot2016sensitivity}, where the effects of two interacting parameters on the quadratic error is discussed also in the context of their consequences to the IMP and the optimisation. In their work, the two interacting parameters are the pre-exponential factor~$A$ and the activation energy~$E$ of the Arrhenius equation, and the quadratic error is calculated over the material mass loss rate. \cite{batiot2016sensitivity} used the Sobol indices to discuss the well-known compensation effect between $A$~and~$E$ in terms of the interaction effects captured by the second-order index. Given the similarity between the applied methodologies and the produced outcomes, it could be said also here that the linear relation between values of specific heat at 150~°C and at 480~°C translate into compensation effects and trade-offs during the IMP.

\begin{figure}[!htp]
     \centering
     \begin{subfigure}[b]{0.49\textwidth}
         \centering
          \includegraphics[width=\textwidth]{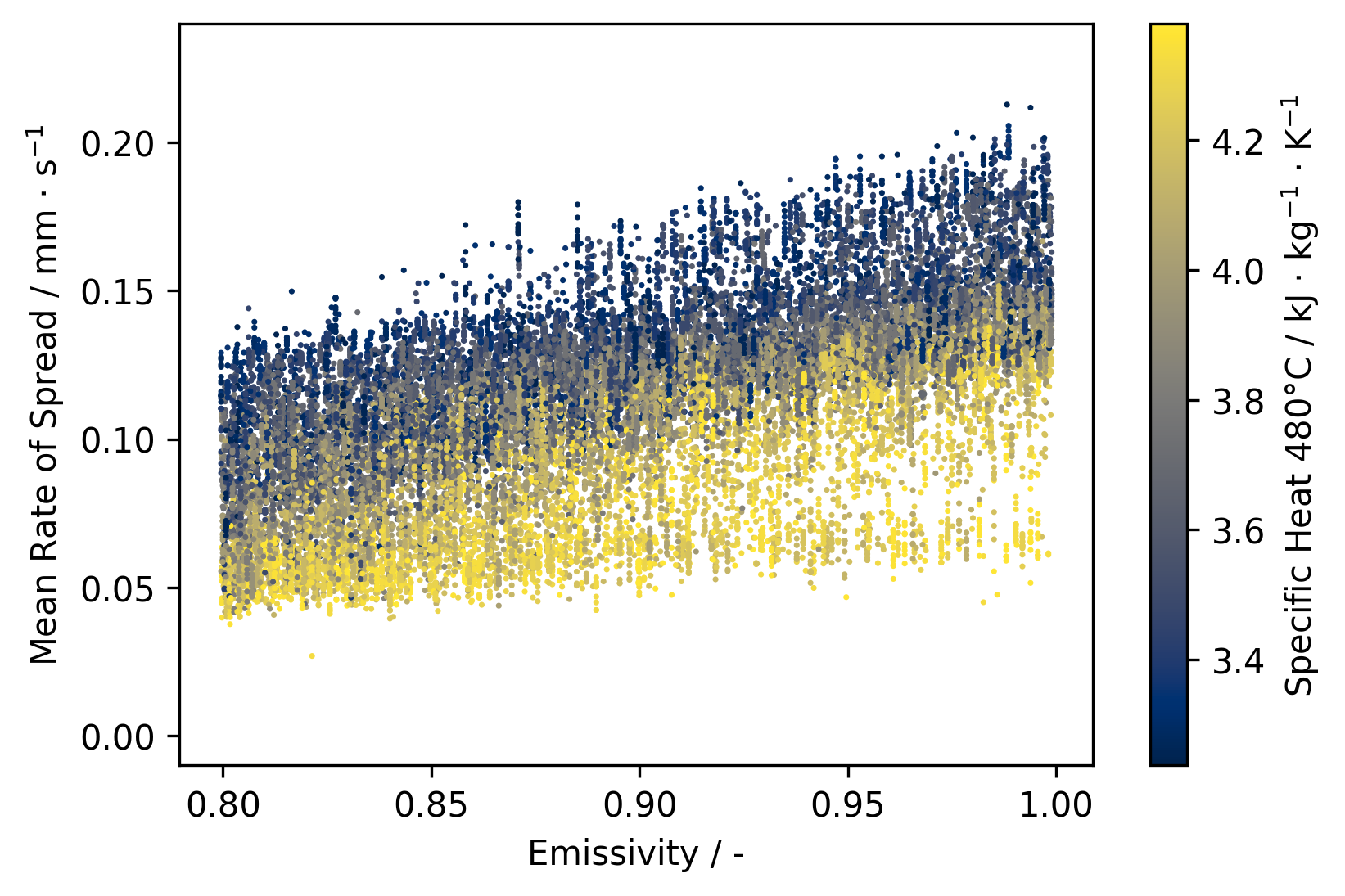}
         \caption{}
         \label{fig-8a}
     \end{subfigure}
     \hfill
     \begin{subfigure}[b]{0.49\textwidth}
         \centering
         \includegraphics[width=\textwidth]{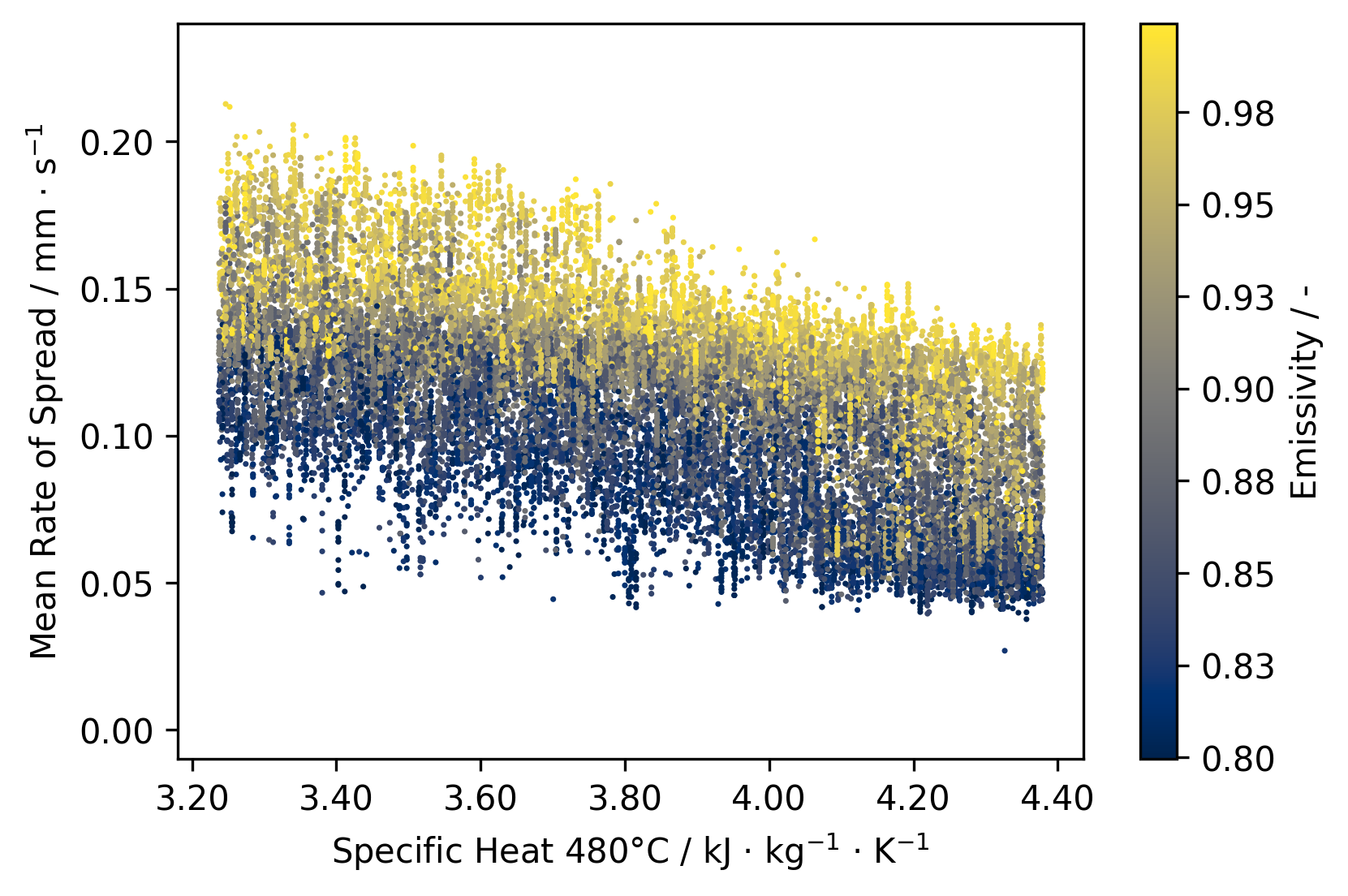}
         \caption{}
         \label{fig-8b}
     \end{subfigure}
        \caption{Scatter plots of (a) MRS plotted against emissivity and coloured by the specific heat at 480~°C; and (b) MRS plotted against specific heat at 480~°C and coloured by emissivity.}
        \label{fig-8}
\end{figure}

The 2-D~scatter plots of the MRS versus emissivity and specific heat at 480~°C are shown respectively in Figures~\ref{fig-8a} and~\ref{fig-8b}. The parameters are the two most influential ones to impact the MRS, according to what is shown in Figure~\ref{sts1mrs}. It can be seen that the relationship between the MRS and each individual parameter can be approximated by a linear function. However, whereas increasing values of emissivity act to increase the MRS, an opposite effect on the MRS is observed when the specific heat at 480~°C is increased, similarly to what was previously discussed in Section~\ref{rmse_mrs}. This behaviour is coherent to the modelling of heat transfer mechanisms. The higher the value of emissivity is, the more heat by radiation is absorbed by the material. More absorbed heat causes local temperatures in the material to rise faster, which then enhances the pyrolysis rates. Increased pyrolysis rates accelerates the production of fuel gases which subsequently burn, releasing more heat which then feeds the positive feedback loop that sustains the spread. 

\section{Conclusions}
\label{sec:conclusions}

The SAs conducted in this study provided meaningful information on the differences in parameter importance to a Cone Calorimeter and a flame spread simulation conducted with FDS. Sobol indices of the input parameters suggested that the Cone Calorimeter simulation is not sufficiently sensitive to all of the parameters that are important to the flame spread. This is an issue because unimportant parameters to the Cone Calorimeter simulation will be estimated with higher degree of uncertainty during the IMP, which is then carried over to the flame spread simulation. In addition, it was revealed that the brief importance of some parameters in the Cone Calorimeter is diminished when the temporal development of the HRR is summarised in a single-value output, as the global RMSE value. Only the values of specific heat at 150~°C and at 480~°C seemed to influence the RMSE through strong interaction effects, whereas the importance of the remaining parameters is negligible. Moreover, the relation between the RMSE and its two most important parameters presented by scatter plots helped to visually identify subsets of the input space that could lead to minimised values of RMSE or convergence to local minima during an IMP. A possible solution for a more effective parameter estimation seems to rely on a combination of RMSE calculated at different phases of the experiment, such that the full set of sensitive parameters is covered. Still, many of the important parameters to the flame spread are practically unimportant throughout the Cone Calorimeter simulation. This limitation could possibly be overcome by IMPs based on bench-scale flame spread experiments, in which the MRS is taken directly as a cost function. 

\section*{Data and Software Availability}

The data is made publicly available:  \url{https://doi.org/10.5281/zenodo.7618897}. (Hint: This link is yet not accessible publicly, but the repository is set up and the DOI is reserved. It will be active, after the manuscript was accepted for publication.) 

\section*{Acknowledgements}

We gratefully acknowledge the computing time granted through the project on the CoBra-system, funded by the German Federal Ministry of Education and Research with the grant number 13N15497. This research was partially funded by the German Federal Ministry of Education and Research with the grant number 13N15497.

\section*{Authorship Contribution Statement}

\textbf{Tássia L.S. Quaresma:} conceptualisation, data curation, formal analysis, investigation, methodology, software, validation, visualisation, writing -- original draft preparation, writing -- review and editing

\textbf{Tristan Hehnen:} conceptualisation, validation, writing -- review and editing

\textbf{Lukas Arnold:} conceptualisation, methodology, project administration, resources, software,  supervision, validation, writing -- review and editing, funding acquisition







\bibliographystyle{unsrt}
\bibliography{bibliography}

\end{document}